\documentclass[letterpaper,english,aps,prl,twocolumn,notitlepage]{revtex4-2}
\usepackage[T1]{fontenc}
\usepackage{graphicx}
\usepackage{soul}
\usepackage{comment}
\usepackage{cancel}
\usepackage{mathdots}

\usepackage{amsmath}
\usepackage{amssymb}

\makeatletter

\pdfpageheight\paperheight
\pdfpagewidth\paperwidth

\usepackage{color}
\usepackage{dsfont}
\usepackage{dsfont}

\usepackage[usenames,dvipsnames,svgnames,table]{xcolor}

\definecolor{violet}{rgb}{0.58, 0.0, 0.83}

\newcommand{\mkaddcomment}[1]{{\bf {\color{violet}{[MK: #1]}}}} 

\makeatother

\begin{document}

\title{Zeno physics of the Ising chain with symmetry-breaking boundary dephasing}

\author{Umar Javed$^{1}$, Riccardo J. Valencia-Tortora$^{2}$, Jamir Marino$^{2}$, Vadim Oganesyan$^{3,4,5}$, Michael Kolodrubetz$^1$}
\affiliation{$^1$Department of Physics, The University of Texas at Dallas, Richardson, Texas 75080, USA}
\affiliation{$^2$Institut f\"{u}r Physik, Johannes Gutenberg-Universit\"{a}t Mainz, D-55099 Mainz, Germany}
\affiliation{$^3$Physics program and Initiative for the Theoretical Sciences, The Graduate Center, CUNY, New York, NY 10016, USA}
\affiliation{$^4$Department of Physics and Astronomy, College of Staten Island, CUNY, Staten Island, NY 10314, USA}
\affiliation{$^5$Center for Computational Quantum Physics, Flatiron Institute, 162 5th Avenue, New York, NY 10010, USA} 

\begin{abstract}

In few-qubit systems, the quantum Zeno effect arises when measurement occurs sufficiently frequently that the spins are unable to relax between measurements. This can compete with Hamiltonian terms, resulting in interesting relaxation processes which depend non-monotonically on the ratio of measurement rate to coherent oscillations. While Zeno physics for a single qubit is well-understood, an interesting open question is how the Zeno effect is modified by coupling the measured spin to a non-trivial bulk. In this work, we study the effect of coupling a one-dimensional transverse field Ising to a Zeno spin which lives at the boundary. We find that sharp singularities occur in the boundary relaxation dynamics, which can be tied to the emergence or destruction of edge modes that can be found analytically. Finally, we provide numerical evidence that the dynamical singularities are stable in the presence of integrability-breaking interactions.

\end{abstract}

\maketitle

\begin{figure}[b]
    \centering
    \includegraphics[width=0.7\columnwidth]{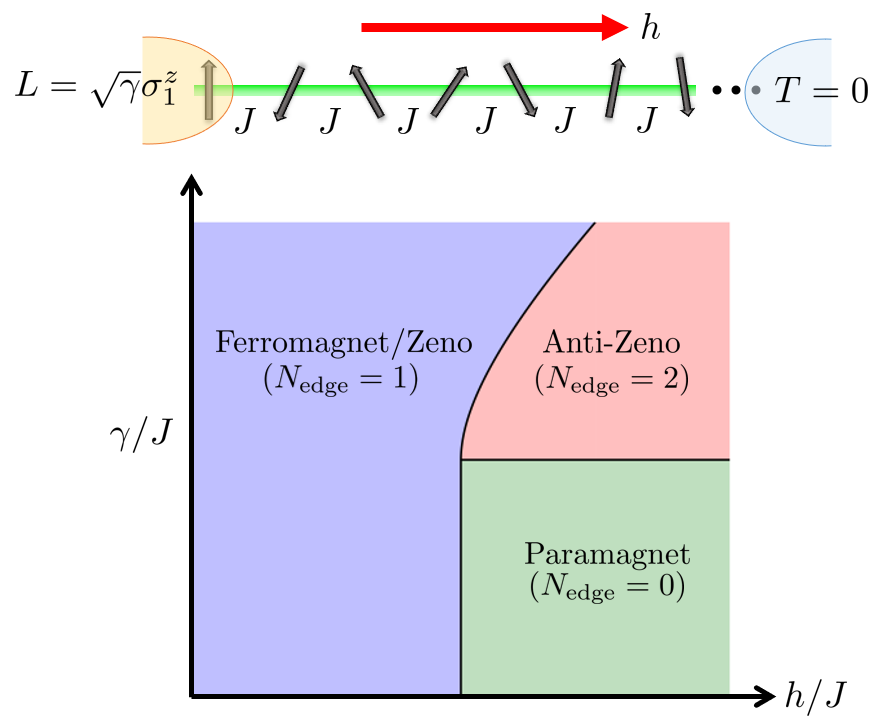}
    \caption{Illustration of Ising model with symmetry-breaking boundary noise. Far from the boundary, the Ising model is in its ground state ($T=0$). Boundary spin dynamics show singularities which can be connected to edge modes, resulting in a nontrivial boundary phase diagram.}
       \label{fig:model}
\end{figure}

Understanding the nature of non-equilibrium correlated phases of matter in many-body systems beyond the perturbative regime is challenging. Yet nonequilibrium quantum physics is precisely the regime that is increasingly experimentally accessible in systems ranging from cold atoms and ions to superconducting circuits to (noisy) quantum computers \cite{Langen2015,Bruzewicz2019,Kjaergaard2020}. Therefore, characterizing nonequilibrium quantum systems and their phases of matter is one of the most important questions in many-body physics today.

Given the dearth of solvable non-equilibrium models, it is crucial to develop an in-depth understanding of the few models in which analytical methods exist. One important class of such systems are quantum boundary/impurity models. Examples where these can be treated beyond the perturbative limit include the Anderson impurity~\cite{nozieres1969singularities,anderson1967infrared,silva2008statistics,cetina2016ultrafast,knap2012time,tonielli2019orthogonality},  transport in one-dimensional junctions with strong electron-electron interactions ~\cite{kane1992transport,kane1995impurity}, and the generation and growth of entanglement among magnetic impurities and their surrounding environments in the Kondo effect ~\cite{latta2011quantum, cox1998exotic, andrei1995}. A key tool for addressing many of these example systems is boundary conformal field theory (CFT), which uses generalized scale invariance near quantum critical points to build a significant analytical toolkit \cite{PhysRevB.96.241113,Vasseur,Francica,Prev_2,Kun,campostrini2014finite,Campostrini}.

However, most results in boundary CFT do not address the increasingly important situation of open quantum systems. Motivated by developments in quantum simulation and computation, it is increasingly important to consider quantum systems which are driven far-from-equilibrium yet, inevitably, remain coupled to their environments. Despite hindering some goals of quantum information theory, dissipation has been found to be useful in other ways such as aiding in state preparation \cite{Budich2015,Reiter2016,Harrington2022} and enhancing the phase structure of quantum matter \cite{Diehl2008,sieberer2016keldysh,Rakovszky2023}. Nonequilibrium open quantum systems are even harder to treat in the strongly correlated regime, necessitating the study of models where concrete predictions can be made.

In this paper, we show that the transverse field Ising chain with symmetry-breaking boundary dephasing enables analytical predictions for the boundary dynamics. Despite symmetry breaking at the boundary, it remains integrable, with certain correlation functions expressible in the language of free Majorana fermions. Edge modes emerge which are naturally described in terms of the Majoranas, giving rise to an interesting (boundary) phase diagram. Furthermore, we show that these edge modes are directly connected to boundary spin dynamics, such that the appearance or disappearance of edge modes coincides with sharp changes in the dynamics. We argue that these singularities are robust to integrability-breaking interactions in the appropriate scaling limit and support the argument with matrix product state-based numerics. Finally, we argue that this is a fundamentally nonequilibrium phenomenon that is hidden from conventional equilibrium observables. We show this explicitly for the steady state energy current from the bath and boundary magnetic susceptibility.

\section{Model}

We consider the one-dimensional Ising model with open boundary conditions and dephasing that breaks Ising symmetry at the boundary. Specifically, we assume non-unitary time evolution of the Lindblad form, 
\begin{equation}
    \frac{d \rho}{dt}=\mathcal{L}[\rho]=-i[H,\rho]+ L \rho L ^{\dagger} -\frac{1}{2}\{L^{\dagger} L ,\rho\},
\end{equation}
with Hamiltonian 
\begin{equation}
    H=-J\sum_{n=1} ^{L-1} \sigma_{n} ^z \sigma_{n+1} ^z - h\sum_{n=1} ^{L} \sigma_n ^x
\end{equation}
and a single Lindblad operator
\begin{equation}
    L=\sqrt{\gamma} \sigma_1^z.
\end{equation}
We use the convention $J=\hbar=1$ throughout.

We are interested in the dynamics of the boundary spin, $\sigma^z_1$, in the presence of this boundary dephasing. One motivation for this is the quantum Zeno effect, which for $J=0$ (single spin) says that the spin dynamics get frozen in the limit $\gamma \gg h$ due to repeated measurements by the environment. By adding interactions $J>0$, we wish to understand the effect of many-body physics -- including quantum phase transitions -- on Zeno physics. Furthermore, the phase transition in the Ising model is a canonical example of a conformal field theory (CFT). By coupling this system to a noisy, relevant boundary perturbation, there is hope to draw a connection between many-body Zeno physics and boundary CFT. Similar ideas have been explored in recent works \cite{berdanier2019universal,froml2019fluctuation,dolgirev2020non}, but addressing much different questions. Specifically, \cite{berdanier2019universal} considered a random boundary drive but only in a strongly driven limit that falls outside of the Lindblad approximation and studied many-body Loschmidt echo. \cite{froml2019fluctuation}  and \cite{dolgirev2020non} studied dissipative impurity models similar to ours, but focused on alternative observables such as transport and non-Gaussian correlations that have direct connection to two-time correlations, which is the more conventional observable used to define Zeno physics. By contrast, we study two-time correlations of a boundary impurity, where we will analytically solve for modifications of the (boundary) Zeno effect.

In the absence of static symmetry-breaking boundary field, the spin itself is unpolarized. 
Therefore, the quantity of interest is the two-time correlation function, which is related to the boundary magnetic susceptibility. Specifically, we seek to find the two-time correlation function in the time-evolved density matrix $\rho(t)$, which is given by using the regression theorem \cite{howard}:
\begin{equation}
    C(t,\Delta t)=\langle \sigma_1 ^z (t+\Delta t) \sigma_1 ^ z (t) \rangle =\mathrm{Tr}\left\{ \left[ e^{\mathcal{L^\dagger} \Delta t} \sigma_1^z \right] \sigma_1^z \rho (t) \right\},
\end{equation}
where the adjoint Liouvillian $\mathcal{L}^\dagger$ generates Heisenberg operator evolution. The initial state is $\rho(0) = |\psi_\mathrm{gs}\rangle \langle \psi_\mathrm{gs} |$, the ground state of the unperturbed Ising chain ($\gamma=0$). Then, at $t=0$, a finite value of $\gamma$ is quenched on. Once we introduce this noise at the edge, we expect quasiparticle excitations to travel ballistically to the other edge of the chain in time $t\approx L/v$. Outside the ballistic front ($x > vt$), the system remains in its ground state. Inside the ballistic front ($x<vt$), the system approaches a new quasi-stationary state in which excitations are continuously added via the boundary dephasing.   We are interested in this quasi-stationary non-equilibrium steady state (NESS) that forms near the boundary for $t + \Delta t < L / v$. In order to equilibrate to the local NESS, $t$ must be chosen to be sufficiently large as well; we use $t=40$ throughout our data. Note that this should be equivalent to the true NESS that forms for a large, zero temperature bath placed sufficiently far from the perturbed boundary (see Figure \ref{fig:model}).

Numerically calculating the two-time autocorrelation function remains challenging in general, but for this particular model it can be efficiently obtained by rewriting in terms of Majorana fermions. We start by a conventional Jordan-Wigner transform on the spin degrees of freedom \cite{sachdev_2011}, 
\begin{align}
    \sigma_j^x =i \eta_{2j-1} \eta_{2j},\ \sigma_j^z = \left(\prod_{n=1} ^{j-1} i \eta_{2n-1} \eta_{2n}\right) \eta_{2j-1}.
\end{align}
The boundary spin is then a single Majorana operator, $\sigma_1^z=\eta_{1}$, and the Hamiltonian can be written as
\begin{equation}
    H=- i h \sum_{j=1} ^{L} \eta_{2j-1} \eta_{2j} - iJ\sum_{j=1}^{L-1}\eta_{2j} \eta_{2j+1}.
\end{equation}
The first step is to evolve the density matrix to $\rho(t)$ and calculate the equal-time correlation function $C(t,0)=\mathrm{Tr}\left[  \sigma_1^z \sigma_1^z \rho (t) \right] = \mathrm{Tr}\left[  \eta_{1}^2 \rho (t) \right]$. Since $\eta_{1}^2=1$, this is always equal to $1$. Crucially, we can also calculate the full Majorana correlation matrix $\mathcal{C}_{ij}(t,0) = \mathrm{Tr}\left[  \eta_{i} \eta_j \rho (t) \right]$ such that $C=\mathcal{C}_{11}$. Two-time correlation functions are then able to be calculated because the adjoint Liouvillian conserves Majorana number:
\begin{equation}
\mathcal{L}^\dagger \left[ \eta_{i} \right] =i[H,\eta_{i}]+\gamma(\eta_{1}\eta_{i}\eta_{1}-\eta_{i}) = \sum_{j=1}^{2L} M_{ij} \eta_j.
\end{equation} 
The matrix $M$ has the form
\begin{equation}
   M=\left(\begin{array}{cccc}
0 & -2h & 0 & 0\\
2h & -2\gamma & -2J & 0\\
0 & 2J & -2\gamma & -2h\\
0 & 0 & 2h & \ddots
\end{array}\right).
\end{equation}
One can then readily show that the 2-time correlation matrix evolves as
\begin{equation}
    \frac{\partial \mathcal{C}(t,\Delta t)}{\partial \Delta t} = M \mathcal{C}(t,\Delta t)
\end{equation}
with equal-time correlations, $\mathcal{C}(t,0)$, as an initial condition. Clearly the eigenmodes of $M$ play a crucial role in understanding the dynamics of the boundary spin; we therefore refer to this as the single-Majorana evolution matrix. More details for how these equations are solved numerically may be found in the supplement \ref{sec:sup_numerics}.

We note briefly that an alternative route to solving the open quantum dynamics exists by vectorizing the density matrix and treating the Lindblad evolution via third quantization. This doubles the effective Hilbert space, but manifestly makes the problem integrable because the boundary dephasing maps to a non-Hermitian Ising term connecting the bra and ket Hilbert spaces. A similar technique was used in recent papers \cite{Shibata2020,Zheng2023}, which studied the related problem of dephasing connected to both ends of a finite system and found a similar phase diagram of the edge modes. While formally identical to our method, third quantization makes it more challenging to connect spectral features and eigenmodes of the Liouvillian to physical observables, so we primarily use the single-particle evolution matrix throughout this paper. More details on the third quantization formalism and its connection to the single-particle evolution matrix $M$ can be found in the supplementary information in sections \ref{sec:supp_third_quantization}  and \ref{sec:supp_connecting_second_and_third_quantization}.

\section{Results}

\begin{figure}
    \centering
    \includegraphics[width=\columnwidth]{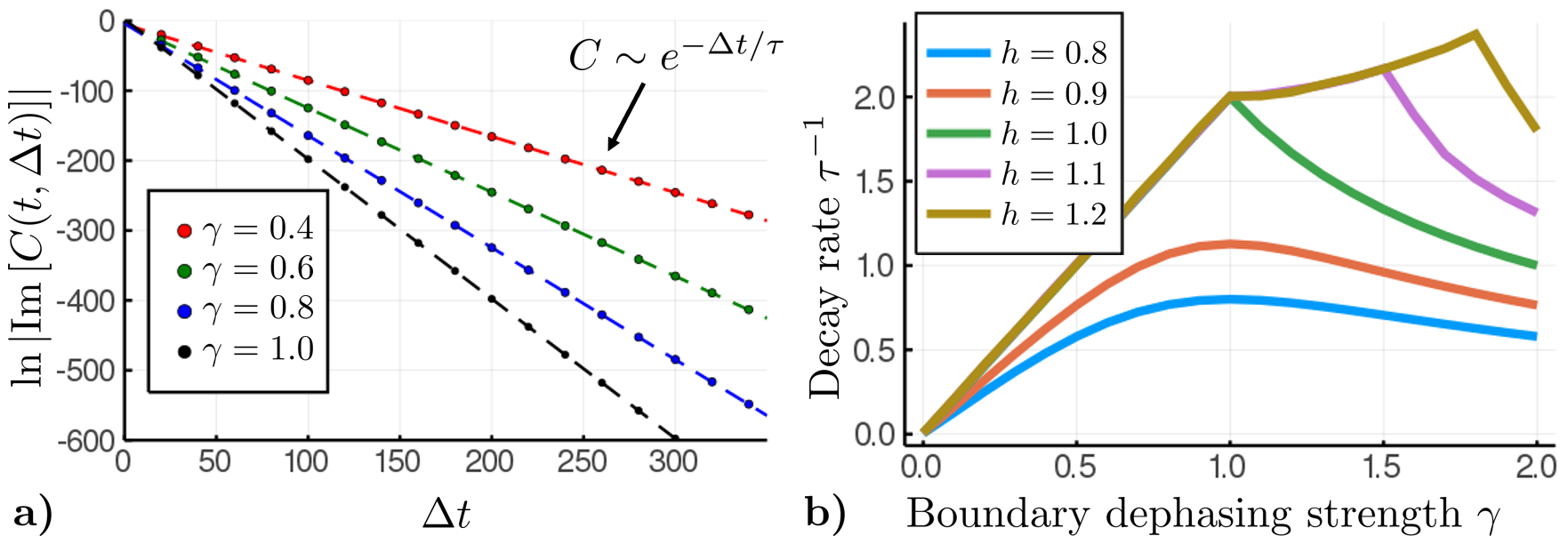}
    \caption{(a) Numerical results for two-time correlation function of the edge spin as a function of boundary dephasing rate ($\gamma$) for $L=1000$, $J=h=1$, and $t=40$. (b) Plot of decay rate $\tau^{-1}$ as a function of $\gamma$ for different values of $h$. }
       \label{fig:correlation_function}
\end{figure}

Using the free fermionic form, we numerically solve for the 2-time correlations of the boundary spin. For all parameters chosen, the (complex-valued) autocorrelation function is found to decay exponentially at late times $\Delta t$ as $C \sim C_0 \exp[-\Delta t/\tau]$, where $\tau$ is the relaxation time scale the characterizes the Zeno effect. 
At short times $\Delta t$, other fast-decaying eigenmodes of $M$ participate in the dynamics, spoiling the simple exponential decay. We fit the numerical data for large $\Delta t$ to extract a value of $\tau$, as shown in Figure \ref{fig:correlation_function} a.

Plotting the relaxation rate $\tau^{-1}$ as a function of $\gamma$ and $h$ (Figure~\ref{fig:correlation_function}b), we find the striking result that, for $h \geq 1$, sharp singularities occur in this decay constant in a fashion reminiscent of equilibrium observables in a conventional first order phase transition. These singularities separate the parameter space into three distinct phases (Figure \ref{fig:model}, bottom), which we label paramagnet (PM), ferromagnet (FM)/Zeno, and anti-Zeno. The PM and FM smoothly connect to the respective ground state phases at $\gamma=0$ with a transition that extends vertically from $h=J$. The anti-Zeno phase only appears at $\gamma > J$ and, as we will see, cannot be thought of as smoothly connected to the ground state Ising physics.

\begin{figure*}
    \centering
    \includegraphics[width=0.85\textwidth]{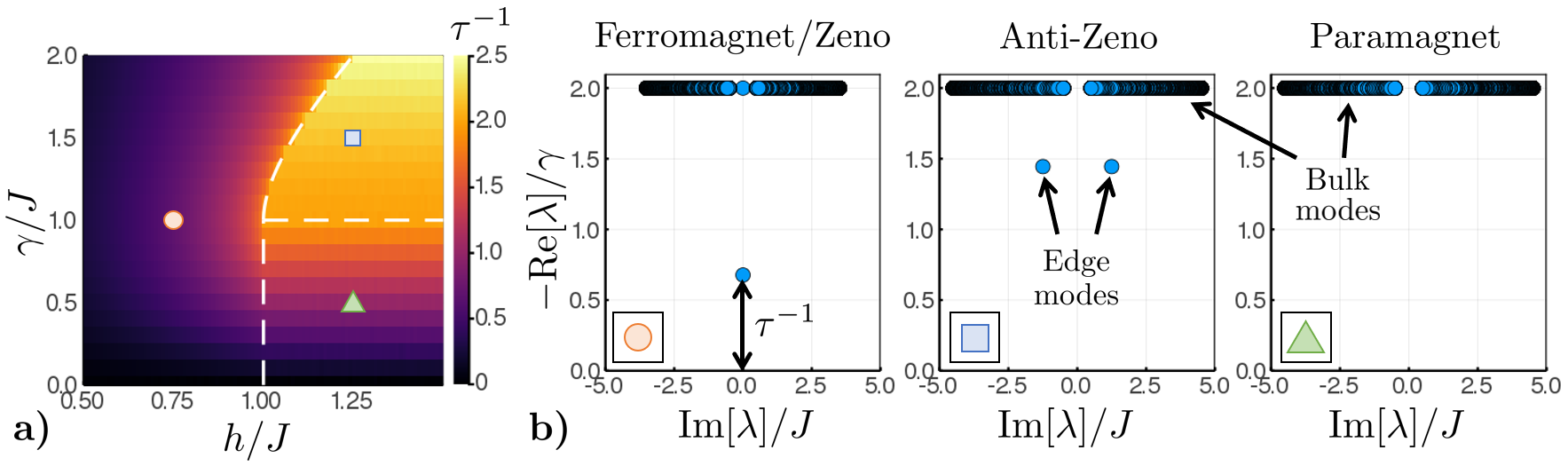}
    \caption{(a) Phase diagram obtained from singularities of the relaxation rate $\tau^{-1}$. (b) Eigenvalues $\lambda$ of the Majorana evolution matrix $M$ plotted for three different points in the phase diagram (left panel), showing number and structure of edge modes. The decay rate is set by the mode with the smallest magnitude real component.}
    \label{fig:spectrum}
\end{figure*}

In order to understand the origin of these singularities, an interesting analogy can be drawn to our previous work \cite{javed2023counting}, which examined the Ising chain with a static symmetry-breaking boundary field. There, we observed a connection between the two-time correlation function of the edge spin and emergent edge modes in the Majorana problem. While this setup is qualitatively different, we are inspired to ask a similar question of the effective non-Hermitian matrix $M$ which determines time evolution of the Majorana correlations. We begin by solving the eigenmodes of $M$ numerically, results of which are shown in Figure \ref{fig:spectrum}b. There are clearly modes that appear outside of the bulk, which we confirm are edge modes localized near the site with dephasing. Furthermore, we confirm that the change in edge mode counting is directly tied to the singularities in $\tau$.

In fact, we can go one step further and solve for the edge modes of $M$ as well as their phase transitions analytically. We consider the following (unnormalized) ansatz for the edge mode
\begin{equation}
\eta_{\text{edge}}=\sum_{j=1}^{N}\left(r^{j-1}\eta_{2j-1}+Ar^{j-1}\eta_{2j}\right),
\end{equation}
where $\Big| r=e^{-1/\xi + i\varphi} \Big| < 1$  describes the decay with length scale $\xi > 0$.  We are interested in the ``eigenenergy'' $\lambda$  such that $M\eta_\mathrm{edge}=\lambda \eta_\mathrm{edge}$.  This equation is solved in the section~\ref{sec:supp_edge_modes_M}, from which we observe that the number of physically meaningful edge modes ($|r| < 1$) changes sharply at phase boundaries given by
\begin{equation}
    h_c = \begin{cases}
        J & \mathrm{for}~\gamma < J \\
        \frac{\gamma^{2}+J^2}{2\gamma} & \mathrm{for}~\gamma > J
    \end{cases}
\end{equation}

Moreover, this analytical solution provides a complete description of the edge modes and their phase transitions, allowing us to re-interpret the observed dynamical signatures physically. For example, we readily see that the PM and FM edge modes track continuously to the Majorana zero modes of the Kitaev chain as $\gamma \to 0$.  As $h$ approaches $J$ from the FM side, the edge mode gap approaches the bulk dephasing rate of $2\gamma$, such that the edge mode merges into the bulk simultaneously with the bulk gap closing at $h=J$. The phase transition looks similar to the conventional ground state Ising transition, with diverging edge correlation length $\xi \to \infty$ and a dissipative gap $\Delta \sim |h-J|$  suggesting that $\nu=z=1$ as in the ground state.

Another relatively simple limit of the model is $\gamma \gg J$, for which the physics of the single boundary spin becomes dominant. In this case, the limits $\gamma \ll h$ and $\gamma \gg h$ correspond to the anti-Zeno and Zeno regimes, respectively. While there is not a conventional phase transition in the steady state of the single-spin Zeno model, there is an exceptional point (i.e. under- to over-damped) transition at $\gamma_c=2h$.  This is precisely the asymptote that we find for our phase boundary when  $\gamma,h \gg J$.  However, all the way down to $\gamma=J$ there is an exceptional point transition in the spectrum of $M$, allowing us to argue that one has Zeno and anti-Zeno ``phases.'' 

The final boundary transition happens at $\gamma=J$ for $h>J$, between the PM and anti-Zeno phases. At this transition, two non-Hermitian bound states emerge out of the PM continuum at finite momentum. The critical behavior appears Ising-like, with imaginary gap that opens linearly with the tuning parameter $\gamma$ ($\nu=z=1$). Furthermore, the entire transition line has diverging correlation length $\xi \to \infty$ (i.e., $|r|\to 1$) but non-zero momentum $\varphi \neq 0$ except at the tricritical point $h=J=\gamma$. While this transition occurs at finite dephasing strength and away from bulk criticality, this diverging length scale suggests that an appropriately defined field theory with gapped bulk may be possible to construct, for which the transition may be universal.

\section{Integrability breaking perturbations}

\begin{figure*}
\centering
\includegraphics[width=0.8\textwidth]{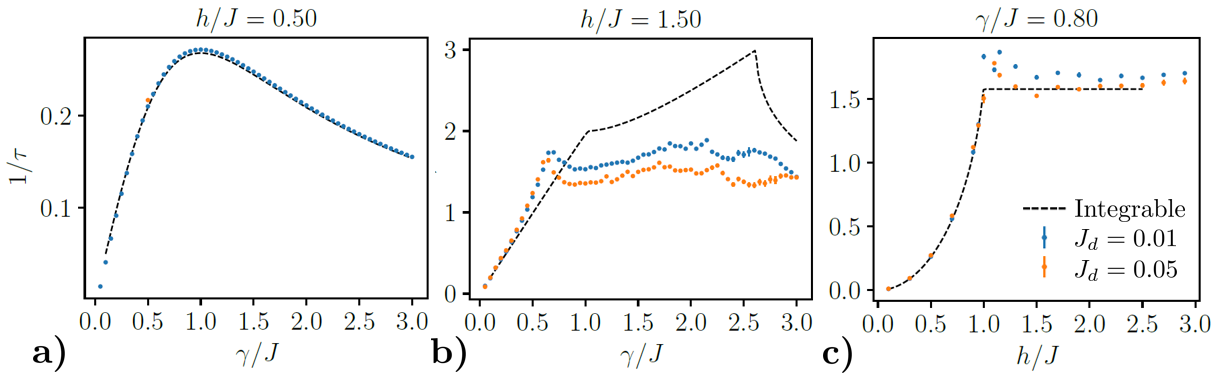}
\caption{Relaxation rate extracted from TEBD simulations of the Ising chain with self-dual interactions for three cuts through the phase diagram: (a) $h/J=0.5$, for which no phase boundary is expected, (b) $h/J=1.5$, for which we expect to go from PM to anti-Zeno and finally to FM/Zeno, and (c) $\gamma/J=0.8$, which should cross the PM/FM phase boundary. The results are given by fitting either a pure exponential $e^{-\Delta t/\tau}$ or one dressed by a power law $(\Delta t)^{-3/2} e^{-\Delta t/\tau}$ in the $h>1$ and $\gamma<1$ region.  The dashed lines in the plots $1/\tau$ are the results from the integrable case. For small $\gamma$, oscillations are still clearly present in $C(t,\Delta t)$ for the small $\Delta t$ available, making the fitting procedures less accurate.}
\label{fig_final_plots_novel_fit}
\end{figure*}

In order to probe universality   of this phase diagram, we now study the robustness of this edge physics in the presence of integrability-breaking interactions. We will do so by modifying the Hamiltonian to 
\begin{equation}
\label{eq_H3}
\begin{split}
H_\mathrm{int} =& -J \sum_{j} \sigma^z_j \sigma^z_{j+1} - h \sum_j \sigma^x_j +\\
&- J_d\sum_{j} \left(\sigma^x_j \sigma^x_{j+1}+\sigma^z_j \sigma^z_{j+2}\right).
\end{split}
\end{equation}
The form of the interaction is chosen such that it is invariant under the Kramers-Wannier (spin/bond) duality, meaning that if a ground state phase transition exists in the bulk, it must occur at the self-dual point $h=J$. 

Since free fermion methods do not work for this interacting model, we must resort to alternative numerical methods. Specifically, we solve the dynamics via a variant of time-evolving block decimation (TEBD) in which the density matrix is treated in third quantization and unfolded such that boundary dephasing becomes a non-Hermitian perturbation at the center of the spin chain (see supplement \ref{sec:supp_third_quantization}). We measure the autocorrelation functions for various $J_d > 0$ (a ferromagnetic perturbation) after first evolving to time $t=1/J$. We consider system size $L=30$, which is sufficient to approximate $L=\infty$ for the timescales we are able to simulate, since no  appreciable finite-size effects occur before $t \sim L/J$ .

We tested our methods by first applying them to the integrable case $J_d = 0$ (see \ref{sec:tebd_non_interacting}) to confirm that it reproduced the known analytical and numerical results from free fermions. One issue that arises in the TEBD numerics is the relatively small $\Delta t \sim 15$ that can be accessed numerically. Specifically, within the PM phase a simple exponential does not fit the data well. The origin of this mismatch comes from the absence of edge modes, meaning that relaxation in the PM comes from the bulk continuum. However, after extracting the exponential prefactor, the remaining power law decay of the boundary spin correlations is strikingly similar to that of the Ising chain with static boundary field \cite{javed2023counting}. Calculating this decay via saddle point approximation gives a $t^{-3/2}$ multiplying the exponential, which is found to give a much cleaner fit and therefore reproduce $\tau^{-1}$ more accurately.

We then apply the same fitting procedure to the values of $C(t,\Delta t)$ obtained in the presence of integrability-breaking interactions. The results (in Figure~\ref{fig_final_plots_novel_fit}) are far from clear, but do show strong signatures of the same phase transitions at both the PM/anti-Zeno and FM/PM transitions. This suggests that the transitions do indeed survive on this time scale despite the presence of integrability breaking. Furthermore, while the data does not resolve a singularity between the FM/Zeno and anti-Zeno phase due to the accessible short times, that limit is expected to be the most robust to interactions because it occurs on finite length scale $\xi$, which becomes particularly small in the limit $h,\gamma\gg J$. We therefore speculate that all the transitions survive interactions on this time scale, though detecting them may be challenging.

At late time $t+\Delta t \gg L / v$, one expects the system to heat up. With just boundary dephasing, the eventual steady state would correspond to infinite temperature. However, there are a few caveats which open the possibility that some of these phenomena will be infinitely long-lived. First, if the chain is semi-infinite, then time $L/v$ can never be reached. This is quite similar to the situation shown in Figure~\ref{fig:model} in the sense that the semi-infinite chain can be thought of as an unbounded zero temperature heat sink. If the mean free path $\ell_\mathrm{mfp}$ is much longer than the edge mode localization length $\xi$, then one would expect these excitations to be carried out of the system by the zero temperature bath before they can backscatter and thermalize the edge mode. The precision of this statement, including whether it turns transitions into crossovers, remains unclear; it will be an important topic for future investigation.

\section{Single-time observables}

Having identified the emergence of edge modes that are tied to the dynamical singularities, we may expect that signatures of these edge modes will also show up in other observables signifying properties of the nonequilibrium steady state. In the supplementary material, we study magnetic susceptibility and energy current in the NESS (sections \ref{sec:supp_magn_susc} and \ref{sec:supp_energy_current}). We find no evidence for singularities of either quantity. This is tied to the evolution of the equal-time correlation matrix $\mathcal{C}(t,\Delta t=0)$, which is governed by a matrix similar to $M$ but with exponential decay of the edge rather than bulk. This reduces the impact of the edge modes in quasi-equilibrium observables and, as we argue in the supplement, appears to prevent singularities in these quantities. Therefore, we conclude that the edge transitions are a dynamical phenomenon which are not captured by  conventional equilibrium properties.

\section{Conclusion}

We have investigated the transverse field Ising chain in the presence of boundary dephasing. We calculated the decay rate of two-time correlations of the boundary spin and showed that sharp singularities occur in between the Zeno and anti-Zeno regimes. The corresponding phase diagram divides into three distinct phases whose boundary relaxation dynamics are directly tied to analytically solvable edge modes in the related Majorana problem. This phase diagram appears robust to interactions in TEBD simulations, consistent with our prediction that the dynamical phenomena will be universal within one-dimensional models with Ising symmetry.

Our results have potentially far-reaching implications if they can be generalized to open versions of other well-studied models, most notably boundary CFTs with dephasing that couples to a relevant boundary perturbations with larger critical exponents. For the Ising CFT with symmetry-breaking boundary magnetic field, the perturbation is relevant near the static Hamiltonian ground state. However, if one attempts to apply similar renormalization group analysis to the symmetry-breaking boundary Lindbladian using Keldysh field theory, one finds that the Lindblad dephasing is marginal. Since the FM/PM transition remains unmodified by weak $\gamma$, we argue that $\gamma$ is marginally irrelevant. By contrast, for the Potts models with $\mathds{Z}_q$  symmetry, a similar Keldysh renormalization group analysis suggests that symmetry-breaking boundary dephasing is relevant for $q\geq 3$ \cite{Cardy1984_bcft_dims}. This would imply the creation of a new boundary phase near the critical point; our work to find this behavior remains on-going. Similar non-unitary boundary terms can be studied for a wide class of models, including experimentally relevant situations such as a dissipative Kondo problem which is known to be equivalent to a boundary CFT \cite{Affleck1995}. Finally, we note that dissipative impurity problems are naturally created in a variety of experimental settings where their dynamics remain challenging to solve. One important example of this is resonant inelastic x-ray scattering (RIXS), in which a localized ``core hole'' impurity modifies the dynamics of the itinerant electrons around it. The problem naturally involves relaxation dynamics as electrons attempt to fill this hole. Therefore, we hope that some of the techniques that we have developed here for an analytically tractable  Ising model with boundary dephasing can be generalized to this more challenging, but experimentally relevant, setting.

\section{Acknowledgments}

We greatly thank Matthew Foster and Romain Vasseur for helpful discussions. This work was performed with support
from the National Science Foundation (NSF) through award numbers MPS-2228725 and DMR-1945529, the Welch foundation through award number AT-2036-202004, and the University of Texas at Dallas Office of Research (M.K. and U.J.).  Part of this work was performed at the Aspen Center for Physics, which is supported by NSF grant No. PHY-1607611. 
The work of RJVT and JM has been supported by the Deutsche Forschungsgemeinschaft (DFG, German Research Foundation) through the grant
HADEQUAM-MA7003/3-1; by the Dynamics and
Topology Center, funded by the State of Rhineland Palatinate.
Parts of this research were conducted using the Mogon supercomputer and/or advisory services offered by Johannes Gutenberg University Mainz (\url{hpc.uni-mainz.de}), which is a member of the AHRP (Alliance for High Performance Computing in Rhineland Palatinate,  \url{www.ahrp.info}), and the Gauss
Alliance e.V. RJVT and JM gratefully acknowledge the computing time granted on the Mogon supercomputer at Johannes Gutenberg University Mainz (\url{hpc.uni-mainz.de}) through the project ``DysQCorr.''
The Flatiron Institute is a division of the Simons Foundation.

\onecolumngrid

\section*{Supplementary information}

In this supplement, we provide additional information and data on
Majorana time evolution, analytical solutions for the edge states,
third-quantized treatments of the dynamics, equilibrium observables,
and computational methods for the integrability-breaking model.

\section{Details regarding second-quantized Majorana time evolution }

Throughout the main text, we primarily solve Majorana correlation
functions using their direct Heisenberg time evolution. In this section,
we provide more details on how this is done numerically. We also solve
for the edge states analytically in a semi-infinite geometry.

\subsection{Numerics for 1- and 2-time correlation functions \label{sec:sup_numerics}}

The time evolution of our system is given by the Lindblad equation

\begin{equation}
\frac{d\rho}{dt}=\mathcal{L}[\rho]=-i[H,\rho]+L\rho L^{\dagger}-\frac{1}{2}\{L^{\dagger}L,\rho\}
\end{equation}
where $\mathcal{L}$ is the Liouvillian super-operator and $L$ is
a Lindblad operator which has the form 
\begin{equation}
L=\sqrt{\gamma}\sigma_{1}^{z}
\end{equation}
Using the Jordan-Wigner transformation, we can write our Hamiltonian
and Lindblad operator in terms of Majorana fermions as: 
\begin{align}
H & =-iJ\sum_{n=1}^{L-1}\eta_{2n}\eta_{2n+1}-ih\sum_{n=1}^{L}\eta_{2n-1}\eta_{2n}\equiv i\eta^{T}A\eta\\
L & =\sqrt{\gamma}\eta_{1}
\end{align}
where 
\begin{equation}
A=\begin{pmatrix}0 & -h/2\\
h/2 & -0 & -J/2\\
 & J/2 & 0 & -h/2\\
 &  & h/2 & 0\\
 &  &  &  & \ddots
\end{pmatrix}
\end{equation}
is an $2L\times 2L$ real antisymmetric matrix and $\eta=(\eta_{1},\eta_{2},\eta_{3},\ldots)^{T}$
is a column vector of Majorana operators.

We are interested in calculating a 2-time correlation function of
the Majoranas starting from initial state $\rho(t=0)$, which is given
by the regression theorem \cite{howard} as 
\begin{equation}
\mathcal{C}_{ij}(t,\Delta t)=\mathrm{Tr}\left[\eta_{i}e^{\mathcal{L}\Delta t}\left(\eta_{j}\rho(t)\right)\right]\label{eq:C_t_Delta_t}
\end{equation}
First, let's calculate the time evolution of the equal-time correlation
function, $\Delta t=0$. Note that the diagonal elements are always
equal to 
\begin{equation}
\mathcal{C}_{jj}\left(t,0\right)=\mathrm{Tr}\left[\eta_{j}^{2}\rho(t)\right]=1.
\end{equation}
The time derivative of the off-diagonal terms ($i\neq j$) is 
\begin{align}
\frac{d\mathcal{C}_{ij}}{dt} & =\text{Tr}\left[\eta_{i}\eta_{j}\frac{d\rho}{dt}\right]\\
 & =\text{Tr}\left[\eta_{i}\eta_{j}\left(-i[H,\rho]+L\rho L^{\dagger}-\rho\right)\right]\\
 & =\text{Tr}\left[\eta_{i}\eta_{j}\left(\sum_{kl}A_{kl}[\eta_{k}\eta_{l},\rho]+\gamma\eta_{1}\rho\eta_{1}-\gamma\rho\right)\right]
\end{align}
Despite appearing to involve 4-point correlation functions of the
Majoranas, we can simplify the above equation to only involve 2-point
correlations by using the anticommutation relations for our Majorana
operators, $\{\eta_{i},\eta_{j}\}=2\delta_{ij}$, as well as cyclicity
of the trace and antisymmetry of $A$ and $\mathcal{C}$: 
\begin{align}
\sum_{kl}A_{kl}\text{Tr}\left(\eta_{i}\eta_{j}[\eta_{k}\eta_{l},\rho]\right) & =\sum_{kl}A_{kl}\text{Tr}\left(\left[\eta_{i}\eta_{j},\eta_{k}\eta_{l}\right]\rho\right)\\
 & =\sum_{kl}A_{kl}\left\langle \eta_{i}\eta_{j}\eta_{k}\eta_{l}-\eta_{k}\eta_{l}\eta_{i}\eta_{j}\right\rangle \\
 & =\sum_{kl}A_{kl}\left\langle \eta_{i}\eta_{j}\eta_{k}\eta_{l}-2\delta_{il}\eta_{k}\eta_{j}+\eta_{k}\eta_{i}\eta_{l}\eta_{j}\right\rangle \\
 & =\sum_{kl}A_{kl}\left\langle \eta_{i}\eta_{j}\eta_{k}\eta_{l}-2\delta_{il}\eta_{k}\eta_{j}+2\delta_{ik}\eta_{l}\eta_{j}-\eta_{i}\eta_{k}\eta_{l}\eta_{j}\right\rangle \\
 & =\sum_{kl}A_{kl}\left\langle \eta_{i}\eta_{j}\eta_{k}\eta_{l}-2\delta_{il}\eta_{k}\eta_{j}+2\delta_{ik}\eta_{l}\eta_{j}-2\delta_{jl}\eta_{i}\eta_{k}+\eta_{i}\eta_{k}\eta_{j}\eta_{l}\right\rangle \\
 & =\sum_{kl}A_{kl}\left\langle -2\delta_{il}\eta_{k}\eta_{j}+2\delta_{ik}\eta_{l}\eta_{j}-2\delta_{jl}\eta_{i}\eta_{k}+2\delta_{jk}\eta_{i}\eta_{l}\right\rangle \\
 & =2\sum_{kl}\left[\delta_{il}\mathcal{C}_{jk}A_{kl}+\delta_{ik}A_{kl}\mathcal{C}_{lj}-\delta_{jl}\mathcal{C}_{ik}A_{kl}-\delta_{jk}A_{kl}\mathcal{C}_{li}\right]\\
 & =4\left(A\mathcal{C}-\mathcal{C}A\right)_{ij}\\
\text{Tr}\left[\eta_{i}\eta_{j}\eta_{1}\rho\eta_{1}\right]-\text{Tr}\left[\eta_{i}\eta_{j}\rho\right] & =\text{Tr}\left[\eta_{1}\eta_{i}\left(2\delta_{1j}-\eta_{1}\eta_{j}\right)\rho\right]-\mathcal{C}_{ij}\\
 & =2\delta_{1j}\mathcal{C}_{1i}-\text{Tr}\left[\eta_{1}\left(2\delta_{1i}-\eta_{1}\eta_{i}\right)\eta_{j}\rho\right]-\mathcal{C}_{ij}\\
 & =2\delta_{1j}\mathcal{C}_{1i}-2\delta_{1i}\mathcal{C}_{1j}\\
 & =-2\left(\delta_{1j}+\delta_{1i}\right)\mathcal{C}_{ij}=-2\left\{ \Delta_{1},\mathcal{C}\right\} _{ij}\\
\implies\frac{d\mathcal{C}}{dt} & =4\left[A,\mathcal{C}\right]-2\gamma\left\{ \Delta_{1},\mathcal{C}\right\} +4\gamma\Delta_{1}\mathcal{C}\Delta_{1}\label{eq:final_eqn_dC_dt}
\end{align}
where 
\begin{equation}
\Delta_{1}=\left(\begin{array}{cccc}
1\\
 & 0\\
 &  & 0\\
 &  &  & \ddots
\end{array}\right)
\end{equation}
projects onto the first row/column. We also used that 
\begin{equation}
\mathcal{C}A=\left(-\mathcal{C}^{T}\right)\left(-A^{T}\right)=\left(A\mathcal{C}\right)^{T}
\end{equation}
The above calculation is a specific example of the general fact that
our Liouvillian has the form of free Majorana time evolution, and
therefore can be efficiently simulated. Since the right hand side
of Eq.~\ref{eq:final_eqn_dC_dt} is linear in $\mathcal{C}$, we
can also vectorize $\mathcal{C}$ and write this as $\dot{\mathcal{C}}=\mathcal{M}\mathcal{C}$,
where $\mathcal{M}$ is a supermatrix. Then the solution is $\mathcal{C}(t)=e^{\mathcal{M}t}\mathcal{C}(0)$.

Having calculated $\mathcal{C}(t,0)$, the 2-time correlation function
is relatively easy, involving similar methods. First note that Eq.~\ref{eq:C_t_Delta_t}
can be thought of as time evolution of $\rho\eta_{j}$ under $\mathcal{L}$
by $\Delta t$, where initial conditions are set by $\mathcal{C}(t,0)$.
A slightly simpler expression is obtained by moving the time evolution
over to $\eta_{i}$ using the adjoint of the Liouvillian, which satisfies
the property that 
\begin{equation}
\text{Tr}\left[A\cdot\mathcal{L}(B)\right]=\text{Tr}\left[\mathcal{L}^{\dagger}(A)\cdot B\right].
\end{equation}
For Hermitian Lindblad operator, as we have here, $\mathcal{L}^{\dagger}$
is identical to $\mathcal{L}$ except with $H\to-H$. Therefore, 
\begin{equation}
\mathcal{C}_{ij}(t,\Delta t)=\mathrm{Tr}\left[e^{\mathcal{L}^{\dagger}\Delta t}\left(\eta_{i}\right)\eta_{j}\rho(t)\right]\implies\frac{\partial\mathcal{C}_{ij}(t,\Delta t)}{\partial\Delta t}=\mathrm{Tr}\left[\mathcal{L}^{\dagger}\left(\eta_{i}\right)\eta_{j}\rho(t)\right].
\end{equation}
Then, via similar machinery as earlier, 
\begin{align}
\mathcal{L}^{\dagger}\eta_{i} & =i[H,\eta_{i}]+\gamma(\eta_{1}\eta_{i}\eta_{1}-\eta_{i})\\
 & =-\sum_{kl}A_{kl}\left[\eta_{k}\eta_{l}\eta_{i}-\eta_{i}\eta_{k}\eta_{l}\right]+\gamma(\eta_{1}\eta_{i}\eta_{1}-\eta_{i})\\
 & =-\sum_{kl}A_{kl}\left[\eta_{k}\left(2\delta_{il}-\eta_{i}\eta_{l}\right)-\eta_{i}\eta_{k}\eta_{l}\right]+\gamma\left(2\delta_{i1}-\eta_{i}\eta_{1}\right)\eta_{1}-\gamma\eta_{i}\\
 & =-\sum_{kl}A_{kl}\left[2\delta_{il}\eta_{k}-\left(2\delta_{ik}-\eta_{i}\eta_{k}\right)\eta_{l}-\eta_{i}\eta_{k}\eta_{l}\right]+\gamma\left(2\delta_{i1}-\eta_{i}\eta_{1}\right)\eta_{1}-\gamma\eta_{i}\\
 & =4\sum_{k}A_{ik}\eta_{k}+2\gamma\eta_{i}\left(\delta_{i1}-1\right)\\
\frac{\partial C(t,\Delta t)}{\partial\Delta t} & =4AC+2\gamma\left(\Delta_{1}-\mathds{1}\right)C\\
\implies C(t,\Delta t) & =\exp\left[4A+2\gamma\left(\Delta_{1}-\mathds{1}\right)\right]C(t,0)\equiv\exp\left[M\Delta t\right]C(t,0)
\end{align}

The above equations give a slightly different perspective on the edge
modes, namely as eigenmodes of $M$. If we say that $v$ is an edge
eigenmode of $M$ with (complex) eigenvalue $\lambda_{v}$, whereas
all the bulk modes are indexed by $k$, then this will clearly show
up as the decay time of the 2-time correlation function because the
evolution with respect to $\Delta t$ follows this same Heisenberg
evolution: 
\begin{equation}
C_{ij}(t,\Delta t)=\mathrm{Tr}\left[\eta_{i}(\Delta t)\eta_{j}\rho(t)\right]\implies\frac{\partial C_{ij}}{\partial\Delta t}=\left(MC\right)_{ij}
\end{equation}
In particular, if $\eta_{i}(0)=\alpha_{v}\eta_{v}+\sum_{k}\alpha_{k}\eta_{k}$,
where $\eta_{v}=v\cdot\eta$ is the edge operator, then 
\begin{align}
\eta_{i}(\Delta t) & =\alpha_{v}\eta_{v}e^{\lambda_{v}\Delta t}\\
 & +\sum_{k}\alpha_{k}\eta_{k}e^{\lambda_{k}\Delta t}\stackrel{\Delta t\to\infty}{\longrightarrow}\alpha_{v}\eta_{v}e^{\lambda_{v}\Delta t}\\
\implies C_{ij}(t,\Delta t) & \stackrel{\Delta t\to\infty}{\longrightarrow}\alpha_{v}C_{ij}(t,0)e^{\lambda_{v}\Delta t}
\end{align}
assuming the edge mode is the slowest-decaying operator.

\subsection{Analytical solution for edge modes \label{sec:supp_edge_modes_M}}

Here we calculate the edge modes of the matrix (with $J=1$)

\begin{equation}
M=\begin{pmatrix}0 & -2h\\
2h & -2\gamma & -2\\
 & 2 & -2\gamma & -2h\\
 &  & 2h & -2\gamma\\
 &  &  &  & \ddots
\end{pmatrix}
\end{equation}
We consider the following (unnormalized) ansatz for the edge mode
\begin{equation}
\eta_{\text{edge}}=\sum_{j=1}^{N}\left(r^{j-1}\eta_{2j-1}+Ar^{j-1}\eta_{2j}\right)=v^{T}\eta,
\end{equation}
where 
\begin{equation}
v^{T}=\begin{pmatrix}1 & A & r & Ar & r^{2} & \dots\end{pmatrix}
\end{equation}
is the vector of coefficients and 
\begin{equation}
\eta^{T}\equiv\begin{pmatrix}\eta_{1} & \eta_{2} & \eta_{3} & \eta_{4} & \eta_{5} & \dots\end{pmatrix}
\end{equation}
is the vector of Majorana operators. Then 
\begin{align}
M\eta_{\text{edge}} & =v^{T}M\eta=Ev^{T}\eta\implies M^{T}v=\lambda v\\
\begin{pmatrix}0 & 2h\\
-2h & -2\gamma & 2\\
 & -2 & -2\gamma & 2h\\
 &  & -2h & -2\gamma\\
 &  &  &  & \ddots
\end{pmatrix}\begin{pmatrix}1\\
A\\
r\\
Ar\\
\vdots
\end{pmatrix} & =\lambda\begin{pmatrix}1\\
A\\
r\\
Ar\\
\vdots
\end{pmatrix}.
\end{align}
This gives three independent equations: 
\begin{align}
2Ah & =\lambda\label{eq:first_eq}\\
-2h-2\gamma A+2r & =\lambda A\label{eq:second_eq}\\
-2A-2\gamma r+2hAr & =\lambda r.\label{eq:third_eq}
\end{align}
Using Equation \ref{eq:first_eq} in \ref{eq:third_eq}, we can cancel
terms and simplify to get 
\begin{align}
A & =-\gamma r\\
\lambda & =2Ah=-2h\gamma r\\
\implies-\cancel{2}h-\cancel{2}\gamma\left(-\gamma r\right)+\cancel{2}r & =\left(-\cancel{2}h\gamma r\right)\left(-\gamma r\right)\\
h\gamma^{2}r^{2}-\left(\gamma^{2}+1\right)r+h & =0\\
r & =\frac{\left(\gamma^{2}+1\right)\pm\sqrt{\left(\gamma^{2}+1\right)^{2}-4h^{2}\gamma^{2}}}{2h\gamma^{2}}
\end{align}

Let's examine a few transitions in this value. For $h=J=1$, we have
\begin{align}
\underline{h=J}:\;r & =\frac{\left(\gamma^{2}+1\right)\pm\sqrt{\left(\gamma^{2}+1\right)^{2}-4\gamma^{2}}}{2\gamma^{2}}\\
 & =\frac{\left(\gamma^{2}+1\right)\pm\left(\gamma^{2}-1\right)}{2\gamma^{2}}\\
 & =\begin{cases}
1\\
\frac{1}{\gamma^{2}}
\end{cases}
\end{align}
The $r=1$ solution corresponds to the $\xi\to\infty$ bulk phase
transition -- and associated edge mode -- which happens for arbitrary
$\gamma$ because the bulk is unaffected by $\gamma$. For $r=1$,
the energy is $\lambda=-2\gamma$, which is precisely the same real
part as the bulk energy, as expected. Meanwhile, for $\gamma>1$,
the additional solution $r=1/\gamma^{2}<1$ exists, which corresponds
to the well-localized edge mode in the FM phase. This has energy $\lambda=-2/\gamma$
which decays slower than the mode at $-2\gamma$, meaning that this
second edge mode is dominant and the additional edge mode that develops
at $h=J$ due to the bulk transition is not seen in late time dynamics.

A second transition happens at $\gamma=J=1$: 
\begin{align}
\underline{\gamma=J}:\;r & =\frac{1\pm\sqrt{1-h^{2}}}{h}
\end{align}
For $h<1$ (FM), this gives one non-trivial solution, corresponding
to the $-$ root; this is again the continuation of the FM edge mode.
For $h>1$, the square root becomes strictly imaginary. Then 
\begin{equation}
\left|r\right|^{2}=\frac{1+\left(h^{2}-1\right)}{h^{2}}=1
\end{equation}
meaning we again have a $\xi\to\infty$ transition, but now with $\text{arg}\left(r\right)\neq0$.
In particular, as $h\to\infty$, we have $r\to\pm i$ suggesting that
the edge mode transition is dominated by bulk modes with $k=\pm\pi/2$.

Finally, we have the exceptional point (a.k.a. over/under-damped)
transition that occurs when the square root in $r$ goes through zero,
causing solutions to $r$ (and therefore $\lambda$) to go from purely
real to complex. This happens when 
\begin{equation}
\left(\gamma^{2}+1\right)^{2}-4h^{2}\gamma^{2}=0\implies h=\frac{\gamma^{2}+1}{2\gamma}.
\end{equation}
Note that this transition only occurs for $\gamma>1$ to satisfy the
requirement that $\left|r\right|\leq1$.

\section{Third quantization}

An alternative route to solving open quantum systems involves promoting
the density matrix to a ``supervector'' and the time evolution function
(Liouvillian) to a non-Hermitian ``superoperator.'' This is commonly
referred to as third quantization. In this section, we discuss how
our model can be solved using third quantization and spell out the
explicit connection to second-quantized Heisenberg time evolution,
as was used in the previous section.

\subsection{Dynamics and edge modes from third quantization \label{sec:supp_third_quantization}}

In third quantization, one begins by vectorizing the density matrix
$\hat{\rho}\to|\rho)$, then turning the Liovillian into a superoperator
$\mathcal{L}\to\mathcal{L}_{\text{op}}$ such that 
\begin{equation}
|\dot{\rho})=\mathcal{L}_{\text{op}}|\rho)\equiv-i\mathcal{H}|\rho).
\end{equation}
In this case, as we'll see, the non-Hermitian Hamiltonian superoperator
$\mathcal{H}$ acts o free Majorana fermions. 

In spin representation, the non-Hermitian Hamiltonian has the ladder
form (see Fig.~\ref{fig:model}b):
\begin{equation}
\mathcal{L}_{\text{op}}=-iH\otimes\noindent\mathds{1}+i\noindent\mathds{1}\otimes H^{T}+\gamma(\sigma_{1}^{z}\otimes\sigma_{1}^{z}-\mathds{1}\otimes\mathds{1)}=-i\mathcal{H}
\end{equation}
Then the Hamiltonian takes the form 
\begin{equation}
\begin{split}\mathcal{H}= & -J\sum_{n=1}^{L-1}\sigma_{n}^{z}\sigma_{n+1}^{z}-h\sum_{n=1}^{L}\sigma_{n}^{x}+J\sum_{n=1}^{L-1}\tau_{n}^{z}\tau_{n+1}^{z}+h\sum_{n=1}^{L}\tau_{n}^{x}+i\gamma\sigma_{1}^{z}\tau_{1}^{z}\end{split}
\end{equation}
where $\sigma$ ($\tau$) acts on the upper (lower) rung. Noting that
the dephasing maps to a complex Ising interaction, this Hamiltonian
can clearly be mapped to a model of free, hopping Majoranas. Specifically,
we unfold the ladder by into a line by with sites labeled $n=-L,-L+1,\cdots,-1$
for the lower ladder and $n=1,2,\cdots,L$ for the upper latter (there
is no site $n=0$). Mathematically, this simply means defining $\sigma_{-n}=\tau_{n}$.
Then we do the same Jordan-Wigner transformation as in the main text,
with the caveat that Majoranas now have indices $\eta_{-2L},\eta_{-2L+1},\cdots,\eta_{-1},\eta_{1},\cdots,\eta_{2L}$
(again label $0$ is skipped). This allows us to write the non-Hermitian
Hamiltonian as 
\begin{equation}
\mathcal{H}=i\eta_{D}^{T}\mathcal{A}\eta_{D}
\end{equation}
where $\eta_{D}^{T}=\left(\eta_{-2L},\eta_{-2L+1},\cdots,\eta_{-1},\eta_{1},\cdots,\eta_{2L}\right)$
includes the full doubled sets of Majoranas and $\mathcal{A}$ is
an antisymmetric matrix of the form
\begin{equation}
\mathcal{A}=\frac{1}{2}\left(\begin{array}{cccccc}
0 & -h\\
h & \ddots\\
 &  & 0 & i\gamma\\
 &  & -i\gamma & 0\\
 &  &  &  & \ddots & h\\
 &  &  &  & -h & 0
\end{array}\right)
\end{equation}

This superoperator matrix determines evolution of the density matrix
$\rho$ in its vectorized form, which shows up directly in the equal-time
correlation function. Similarly, Heisenberg evolution of the Majorana
operators can be obtained from the adjoint Liouvillian, governed by
the same type of effective Hamiltonian with single-particle matrix
$\mathcal{A}^{\dagger}$ (set by $H\to-H$). Therefore, all evolution
properties -- equal-time correlations, two-time correlations, edge
modes, etc -- can equally well be obtained from the third-quantized
formalism. However, it comes at the cost of a larger matrix which
was created by doubling the degrees of freedom. This can make solving
and interpreting the third-quantized data challenging, hence our preference
to consider second-quantized Heisenberg evolution in the main text.

\subsection{Relationship between third-quantized Hamiltonian and second-quantized
Heisenberg evolution \label{sec:supp_connecting_second_and_third_quantization}}

To support our claim that second- and third-quantized formalisms are
equivalent, we now show they are directly related by a unitary basis
transformation. Specifically, consider the second-quantized single-particle
evolution matrix 
\begin{align}
M & =\left(\begin{array}{cccc}
0 & -2h & 0 & 0\\
2h & -2\gamma & -2J & 0\\
0 & 2J & -2\gamma & -2h\\
0 & 0 & 2h & \ddots
\end{array}\right)\equiv M^{\prime}-2\gamma\mathds{1}\\
\text{where}\;M^{\prime} & =\left(\begin{array}{cccc}
2\gamma & -2h & 0 & 0\\
2h & 0 & -2J & 0\\
0 & 2J & 0 & -2h\\
0 & 0 & 2h & \ddots
\end{array}\right)\label{eq:definition_shifted_M_prime}
\end{align}
and the third-quantized single-particle Hamiltonian matrix
\begin{equation}
\mathcal{A}=\frac{1}{2}\left(\begin{array}{cccccc}
0 & -h\\
h & \ddots\\
 &  & 0 & i\gamma\\
 &  & -i\gamma & 0\\
 &  &  &  & \ddots & h\\
 &  &  &  & -h & 0
\end{array}\right).
\end{equation}
Up to a multiplicative prefactor, these matrices (specifically $\mathcal{A}$
and $M^{\prime}$) are clearly quite similar. The difference comes
in the dephasing $\gamma$ term. Therefore, let us begin by trying
to diagonalize the $\gamma$ term in $\mathcal{A}$. This suggests
a unitary transformation
\begin{align}
\mathcal{V} & =\frac{1}{\sqrt{2}}\left(\begin{array}{cccccccccc}
1 &  &  &  &  &  &  &  &  & i\\
 & 1 &  &  &  &  &  &  & i\\
 &  & 1 &  &  &  &  & i\\
 &  &  & \ddots &  &  & \iddots\\
 &  &  &  & 1 & i\\
 &  &  &  & i & 1\\
 &  &  & \iddots &  &  & \ddots\\
 &  & i &  &  &  &  & 1\\
 & i &  &  &  &  &  &  & 1\\
i &  &  &  &  &  &  &  &  & 1
\end{array}\right)\equiv\frac{1}{\sqrt{2}}\left(\mathds{1}+i\mathcal{R}\right)\\
\mathcal{A}_{V} & =\mathcal{V}^{\dagger}\mathcal{A}\mathcal{V}\\
 & =\frac{1}{2}\left(\mathcal{A}+i\left[\mathcal{A},\mathcal{R}\right]+\mathcal{R}\mathcal{A}\mathcal{R}\right)\\
 & =\frac{1}{4}\left(\begin{array}{cccccc}
0 & -2h\\
2h & \ddots\\
 &  & -2\gamma & 0\\
 &  & 0 & 2\gamma\\
 &  &  &  & \ddots & 2h\\
 &  &  &  & -2h & 0
\end{array}\right)
\end{align}
which rotates the bra/ket Majoranas into their $y$-basis: $\eta_{j}\pm i\eta_{-j}$.
This unitary has block-diagonalized the matrix. In fact, the lower-right
block is equal to $-\tilde{M}/4$, where $\tilde{M}$ is the matrix
that determines equal-time correlations (defined formally in Eq.~\ref{eq:M_tilde_def}).
By inverting the rows and columns of the upper left block via further
unitary matrix $\mathcal{R}_{U}$, we can relate that block to the
non-equal-time generating matrix $M^{\prime}=M+2\gamma\mathds{1}$:
\begin{align}
\mathcal{R}_{U} & =\left(\begin{array}{cccccc}
0 & 0 & 1\\
0 & \iddots & 0\\
1 & 0 & 0\\
 &  &  & 1 & 0 & 0\\
 &  &  & 0 & \ddots & 0\\
 &  &  & 0 & 0 & 1
\end{array}\right)\\
\mathcal{A}_{f} & =\mathcal{R}_{U}\mathcal{A}_{V}\mathcal{R}_{U}\\
 & =\frac{1}{2}\left(\begin{array}{cccccccc}
-\gamma & h\\
-h & \ddots & J\\
 & -J & 0 & h\\
 &  & -h & 0\\
 &  &  &  & \gamma & h\\
 &  &  &  & -h & \ddots & J\\
 &  &  &  &  & -J & 0 & h\\
 &  &  &  &  &  & -h & 0
\end{array}\right)=\left(\begin{array}{cc}
-M^{\prime}/4 & 0\\
0 & -\tilde{M}/4
\end{array}\right).
\end{align}
Therefore, we conclude that second quantization and third quantization
yield identical properties and are both solvable using free Majorana
fermions.

\section{Single-time observables}

In this section we treat two quasi-equilibrium observables in the
non-equilibrium steady state: energy current from the bath to the
system and magnetic susceptibility. We show numerically and analytically
that neither has measurable singularities at the locations where edge
state phase transitions occur in the autocorrelation function.

\subsection{Energy Current \label{sec:supp_energy_current}}

The edge state and two-time correlations are properties of the non-equilibrium
steady state obtained by coupling the edge to a particular infinite-temperature
bath and having a zero temperature reservoir deep in the bulk. Thinking
of this as a far-from-equilibrium transport setup, we can ask about
transport of energy, which is the only conserved quantity. In order
to calculate this, noted that the energy taken from bath is equal
to energy absorbed by system. So, we have energy current
\begin{align}
I_{E} & =\frac{d\langle H\rangle}{dt}\\
 & =\frac{d\text{Tr}\left(\rho H\right)}{dt}\\
 & =\text{Tr}\left(\dot{\rho}H\right)\\
 & =\text{Tr}\left[\left(L\rho L^{\dagger}-\frac{1}{2}\left\{ L^{\dagger}L,\rho\right\} \right)H\right]
\end{align}
Note that the Hamiltonian part of time evolution conserves energy,
so drops out. For our Lindbladian, this is 
\begin{align}
I_{E} & =\gamma\text{Tr}\left[\left(\sigma_{1}^{z}\rho\sigma_{1}^{z}-\rho\right)H\right]\\
 & =\gamma\text{Tr}\left[\rho\left(\sigma_{1}^{z}H\sigma_{1}^{z}-H\right)\right]\\
 & =\gamma\text{Tr}\left[\rho\left(-2h\sigma_{1}^{x}\right)\right]\\
 & =-2h\gamma\langle\sigma_{1}^{x}\rangle.\label{IE}
\end{align}
In other words, energy current is proportional to the static transverse
magnetization at the boundary, which is readily calculated using the
equal-time correlation matrix, $\langle\sigma_{1}^{x}(t)\rangle=-i\mathcal{C}_{12}(t)$
. 

\begin{figure}
\centering \includegraphics[width=1\columnwidth]{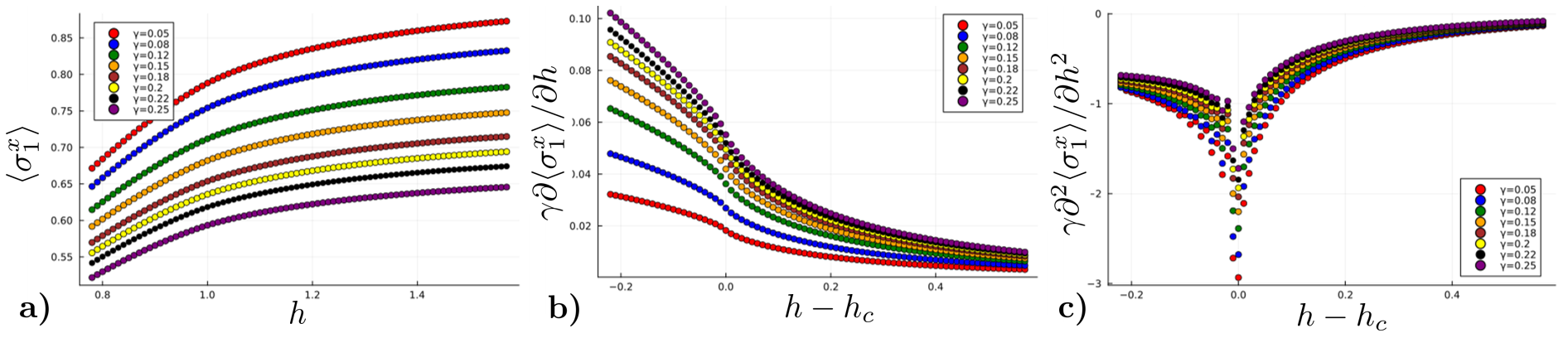}
\caption{Transverse boundary magnetization (which is proportional to energy current
Eq.~\ref{IE}) and its derivatives for different values of $\gamma$ and $h$. For the ground state at $\gamma = 0$, a $|h-h_c|^{-1}$ singularity occurs in the second derivation. Finite $\gamma$ seems to round off this singularity.}
\label{fig:energycurrent} 
\end{figure}

We first study this quantity numerically in the NESS. As shown in
Fig.~\ref{fig:energycurrent}, we don't see the singular behavior
in energy current, which is quite unexpected based on our initial
intuition that there are emergent edge modes at the dissipative boundary
that seem like they should affect the energy current.

To resolve this mystery, let us try solving the equal-time correlation
function in a similar manner to how we did the two-time correlation
function. Starting from Eq.~\ref{eq:final_eqn_dC_dt}, we will now
show how to rewrite it in terms of a single-particle matrix
\begin{equation}
\tilde{M}=4A-2\gamma\Delta_{1}=\left(\begin{array}{cccc}
-2\gamma & -2h & 0 & 0\\
2h & 0 & -2J & 0\\
0 & 2J & 0 & -2h\\
0 & 0 & 2h & \ddots
\end{array}\right). \label{eq:M_tilde_def}
\end{equation}
First, note that 
\begin{align}
\tilde{M}\mathcal{C}+\mathcal{C}\tilde{M}^{\dagger} & =\left[4A-2\gamma\Delta_{1}\right]\mathcal{C}+\mathcal{C}\left[-4A-2\gamma\Delta_{1}\right]\\
 & =4\left[A,\mathcal{C}\right]-2\gamma\left\{ \Delta_{1},\mathcal{C}\right\} .
\end{align}
Next, note that the diagonal elements of $\mathcal{C}$ always equal
to the identity matrix because $\eta_{j}^{2}=1$. The term $4\gamma\Delta_{1}\mathcal{C}\Delta_{1}$
in Eq.~\ref{eq:final_eqn_dC_dt} only affects diagonal elements to
maintain this constraint. Therefore, for the off-diagonal elements
$\mathcal{C}_{\text{od}}$,
\begin{equation}
\frac{d\mathcal{C}_{\text{od}}}{dt}=4\left[A,\mathcal{C}_{\text{od}}\right]-2\gamma\left\{ \Delta_{1},\mathcal{C}_{\text{od}}\right\} =\tilde{M}\mathcal{C}_{\text{od}}+\mathcal{C}_{\text{od}}\tilde{M}^{\dagger}.
\end{equation}

This means that the eigensystem of $\tilde{M}$ determines evolution
of the equal-time correlations. If $\tilde{M}$ has a complete set
of right eigenvectors $v_{j}$ with non-degenerate eigenvalues $\lambda_{j}$
(the special case of exceptional points can be considered separately
if treating the phase transitions), then $v_{j}v_{k}^{\dagger}$ is
an eigenmode of the dynamics:
\begin{align}
\frac{d}{dt}\left(v_{j}v_{k}^{\dagger}\right) & =\tilde{M}v_{j}v_{k}^{\dagger}+v_{j}v_{k}^{\dagger}\tilde{M}{}^{\dagger}\\
 & =\lambda_{j}v_{j}v_{k}^{\dagger}+\lambda_{k}^{\ast}v_{j}v_{k}^{\dagger}-4\gamma v_{j}v_{k}^{\dagger}
\end{align}
We can then use this to calculate the full time evolution. First,
note that $C_{\text{od}}$ is strictly imaginary and antisymmetric,
while $\tilde{M}^{\prime}$ is real but not symmetric. Therefore,
$\tilde{M}^{\dagger}=\tilde{M}^{T}$ implying that $\tilde{M}$ and
$\tilde{M}^{\dagger}$ are isospectral and each eigenvalue $\lambda_{j}$
has a partner $\lambda_{j}^{\ast}$: 
\begin{equation}
\tilde{M}v_{j}=\lambda_{j}v_{j}\implies v_{j}^{\dagger}\tilde{M}^{\dagger}=v_{j}^{\dagger}\tilde{M}^{T}=\lambda_{j}^{\ast}v_{j}^{\dagger}\implies\tilde{M}v_{j}^{\ast}=\lambda_{j}^{\ast}v_{j}^{\ast}.
\end{equation}
For each right eigenvector $v_{j}$, there is also a left eigenvector
$w_{j}^{T}$ such that 
\begin{equation}
w_{j}^{T}\tilde{M}=\lambda_{j}w_{j}^{T}
\end{equation}
A key property is that, while the right eigenvectors $v_{j}$ are
not mutually orthogonal, they do form a completeness relation with
the left eigenvectors:
\begin{equation}
w_{j}^{T}v_{j^{\prime}}=\delta_{j,j^{\prime}}\Leftrightarrow\sum_{j}v_{j}w_{j}^{T}=\mathds{1}
\end{equation}
This allows us to easily decompose the initial correlation function,
$\mathcal{C}_{\text{od}}(t=0)$, into eigencomponents:
\begin{align}
\mathcal{C}_{\text{od},0} & =\sum_{j,k}c_{jk}v_{j}v_{k}^{\dagger}\\
w_{j}^{T}\mathcal{C}_{\text{od},0}w_{k}^{\ast} & =\sum_{j^{\prime},k^{\prime}}c_{j^{\prime}k^{\prime}}\underbrace{w_{j}^{T}v_{j^{\prime}}}_{\delta_{j,j^{\prime}}}\underbrace{v_{k^{\prime}}^{\dagger}w_{k}^{\ast}}_{\delta_{k,k^{\prime}}}=c_{jk}
\end{align}
Finally, we time evolve to obtain
\begin{equation}
C_{\text{od}}(t)=\sum_{j,k}c_{jk}v_{j}v_{k}^{\dagger}\exp\left[\left(\lambda_{j}+\lambda_{k}^{\ast}\right)t\right]
\end{equation}
Like the single Majorana time evolution, this will be dominated by
the slowest-decaying modes, namely those with maximum real part of
$\lambda_{j}$. 

\begin{figure}
\centering \includegraphics[width=0.9\linewidth]{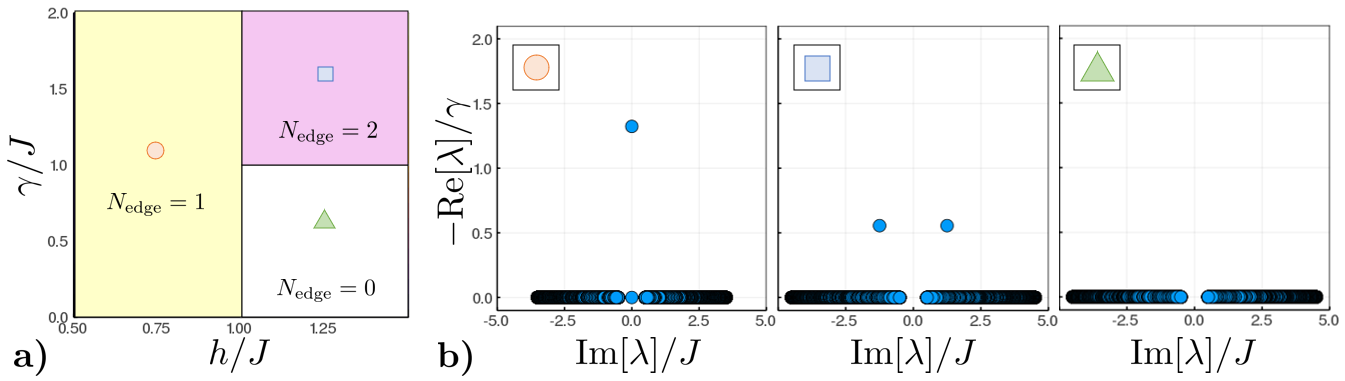}
\caption{(a) Phase diagram showing number of edge modes for equal-time correlation matrix $\tilde{M}$
(Eq.~\ref{eq:M_tilde_def}). Note that counting would be the same for $M$, as here we do not restrict to
modes that are degenerate at slowest decay. (b) Eigenvalues $\lambda$ of $\tilde M$ plotted for three different points in the phase diagram, 
showing that edge modes are fastest-decaying modes for equal-time correlations.}
\label{fig:supp_spectrum_equal_time_corr} 
\end{figure}

The equal-time evolution matrix $\tilde{M}$ looks very similar to
the 2-time evolution matrix $M^{\prime}$ (Eq.~\ref{eq:definition_shifted_M_prime})
with the crucial difference that $\gamma\to-\gamma$. Therefore, when
we go to solve for the edge states, the spectrum looks essentially
identical except that now the edge mode is the fastest-decaying mode,
rather than slowest-decaying. Indeed, due to symmetry of the Hamiltonian
under $h\to-h$ and $J\to-J$, the spectrum of $\tilde{M}$ is equal
to that of $M$ up to multiplication by overall $-1$ and a constant
energy shift, as illustrated in Figure \ref{fig:supp_spectrum_equal_time_corr}.
Because the edge modes are now the fastest decaying one, contrary
to what occurs for the 2-time correlation function, we do not expect
that such edges have an impact in the long-time dynamics of the magnetization.
As a consequence, we do expect singularities. 

\subsection{Magnetic susceptibility \label{sec:supp_magn_susc}}

Having established that singularities only appear in non-equal-time
correlations of the boundary spin, the other observable we consider
is boundary magnetic susceptibility $\chi_{B}$, which is related
to the 2-time correlation function via the Kubo formula. However,
the components occupying these edge states come from equal-time correlations,
where edge mode occupation is suppressed as we saw in the previous
section. To understand which of these is most relevant, we now solve
for magnetic susceptibility numerically.

In order to do so, it is convenient to introduce an additional weak
symmetry-breaking magnetic field $h_{B}$:
\begin{equation}
H\to H-h_{B}\sigma_{1}^{z}.
\end{equation}
The (zero-frequency) magnetic susceptibility is then
\begin{equation}
\chi_{B}=\lim_{h_{B}\to0}\frac{\langle\sigma_{1}^{z}\rangle}{h_{B}},
\end{equation}
where care must be taken to evaluate this in the same NESS as before.
The protocol we choose is as follows:
\begin{enumerate}
\item Initialize the system in its ground state at fixed $(h,J,h_{B})$.
\item Quench on $\gamma>0$ and time evolve to $t<L/v$.
\item Plot $\langle\sigma_{1}^{z}\rangle$ as a function of $t$ to obtain
the NESS magnetization.
\end{enumerate}
In fact, we can do so efficiently using similar free-fermion numerics
by a modified Jordan-Wigner transformation in which an ancillary Majorana
$\eta_{0}$ is introduced, as in \cite{javed2023counting}. The modified
mapping is
\begin{equation}
\sigma_{j}^{x}=i\eta_{2j-1}\eta_{2j},\ \sigma_{j}^{z}=i\eta_{0}\left(\prod_{n=1}^{j-1}i\eta_{2n-1}\eta_{2n}\right)\eta_{2j-1}.
\end{equation}
While this appears to add interactions to the problem, a similar calculation
as before shows that equal-time correlations of the extended Majorana
operator vector, $\eta^{T}\equiv\begin{pmatrix}\eta_{0} & \eta_{1} & \eta_{2} & \dots\end{pmatrix}$,
are governed by the following single-particle evolution matrices:
\begin{align}
\frac{\partial\mathcal{C}(t)}{\partial t} & =4\left[A_{\text{ext}},\mathcal{C}\right]+\gamma\left[4\left(\Delta_{1}\mathcal{C}\Delta_{0}+\Delta_{0}\mathcal{C}\Delta_{1}\right)-2\left\{ \Delta_{0}+\Delta_{1},\mathcal{C}\right\} +4\left(\Delta_{0}+\Delta_{1}\right)\right],
\end{align}
where $\Delta_{j}$ is a diagonal matrix projecting onto the $j$th
component and the extended Hamiltonian matrix is
\begin{equation}
A_{\text{ext}}=\begin{pmatrix}0 & -h_{B}/2\\
h_{B}/2 & 0 & -h/2\\
 & h/2 & 0 & -J/2\\
 &  & J/2 & 0 & -h/2\\
 &  &  & h/2 & 0\\
 &  &  &  &  & \ddots
\end{pmatrix}.
\end{equation}
Then the boundary magnetization is $\langle\sigma_{1}^{z}(t)\rangle=-i\mathcal{C}_{01}(t)$. 

\begin{figure}
\centering \includegraphics[width=0.85\columnwidth]{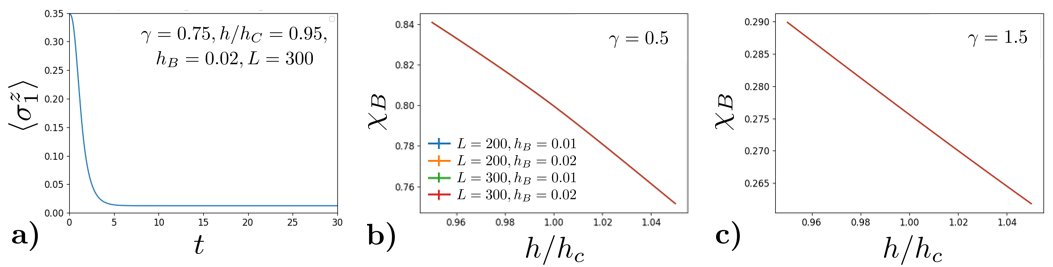}
\caption{(a) Example of data for longitudinal boundary magnetization, $\langle \sigma^z_1 \rangle$, as a function
of time $t$ after a quench from the ground state to finite $\gamma$. (b,c) Boundary magnetic susceptibility obtained
from numerical non-equilibrium steady state ($1 \ll t < L/v$). Curves various $L$ and $h_B$ are nearly identical.}
\label{fig:boundary_susceptibility} 
\end{figure}

Some examples of the boundary magnetization and susceptibility are
shown in Figure \ref{fig:boundary_susceptibility}. Clearly, the system quickly approaches
a boundary steady state. However, it is also apparent from lack of
finite-size or finite-$h_{B}$ effects near the boundary phase transitions
that no low-order singularities exist in the boundary susceptibility.
Combined with the previous section, this provides strong evidence
that the boundary phase transitions are strictly dynamical phenomena.

\section{Additional data for interacting model}

In this section, we provide additional information and data on the
interacting Ising chain and simulations using a variant of time-evolving
block decimation (TEBD). The method proceeds by using third quantization
(see Section \ref{sec:supp_third_quantization}) and unfolding the
vectorized density matrix to create a one-dimensional problem with
dephasing as an impurity in the middle of the chain. Initialization
is done by using DMRG to obtain the ground state of $H$, which is
then used to initial both the left half of the chain (sites $-L$
to $-1$) and the right half (sites $1$ to $L$). Finally, we quench
on the dephasing, which behaves as a non-Hermitian $ZZ$-interaction
in the middle of the chain. Finally, the dynamics is solved via conventional
TEBD techniques. Below, we provide further analysis of our results.

\subsection{Numerical fits for TEBD of non-interacting data \label{sec:tebd_non_interacting}}

\begin{figure}
\centering \includegraphics[width=0.6\columnwidth]{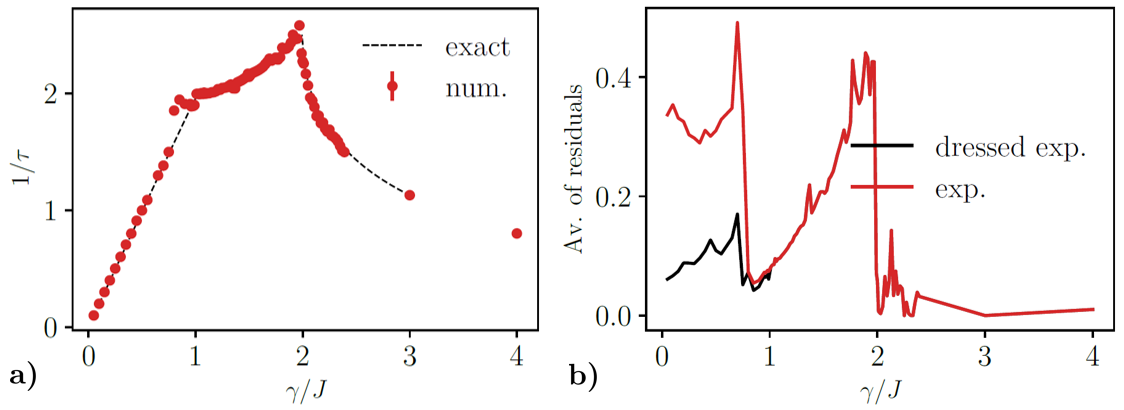}
\caption{Illustration of improved fitting procedure for 2-time correlation function, applied to free 
fermions ($J_d=0$). (a) Time scale $\tau$ obtained from TEBD simulations of $J_d=0$ using the modified
fit procedure with power law prefactor $\sim t^{-3/2}$ for data with $h>J$ and $\gamma < J$. (b) Comparison of
fit residuals between exponential and exponential with power-law prefactor (``dressed'').}
\label{fig:free_fermion_fit} 
\end{figure}

For the non-interacting case ($J_{d}=0$), we know the exact solution
and were able to obtain numerical values of $\tau$ from fits of $C(t,\Delta t)$.
Therefore, it was surprising when, at first, we were unable to obtain
similar quality data for $\tau$ upon fitting $C$ from TEBD. Specifically,
as seen in Figure \ref{fig:free_fermion_fit}, data fitted in the PM phase showed
consistent deviations which became stronger upon approaching the critical
point. 

The solution came from realizing that the analytical solution includes
no edge modes in this phase, meaning that $C$ is solely given by
the bulk contribution
\begin{equation}
C\sim\int dkc_{k}e^{\lambda_{k}t}
\end{equation}
where $c_{k}$ are matrix elements between the initial state and eigenstates
of the $M$ matrix and $\lambda_{k}$ are its eigenvalues correspond
to the bulk mode with momentum $k$. In the bulk, we have that
\begin{equation}
\lambda_{k}=-2\gamma-i\epsilon_{k}
\end{equation}
where $\epsilon_{k}$ is the single-particle dispersion. Factoring
out the global decay rate $e^{-2\gamma t}$, we are left with an integral
of the form $\int dkc_{k}e^{-i\epsilon_{k}t}$. A similar integral
is encountered in solving the boundary spin dynamics of the Ising
chain with static boundary field $h_{B}\sigma_{1}^{z}$ \cite{javed2023counting}.
There, we found that a saddle point approximation gives asymptotic
decay $\sim t^{-3/2}$ at late times. Applying the same logic to the
case with a bulk that is gapped on both the real and imaginary axis,
we obtain late time decay $\sim t^{-3/2}e^{-t/\tau}$.

Re-fitting the data using this power law prefactor significantly improves
the results, as seen in Figure \ref{fig:free_fermion_fit}; deviations remain near the critical
point, where the power law is expected to change when the bulk goes
gapless. Furthermore, the argument of faster decay due to a gapped
bulk density of states should also hold for the interacting model
at low energy. Therefore, we perform this power-law-dressed fit within
the PM phase of the interacting model as well, finding that it decreases
the appearance of unphysical jumps in $\tau$. This is the data shown
in the main text.

\subsection{Alternative integrability-breaking interactions}

\begin{figure}
\centering 
\includegraphics[width=0.95\linewidth]{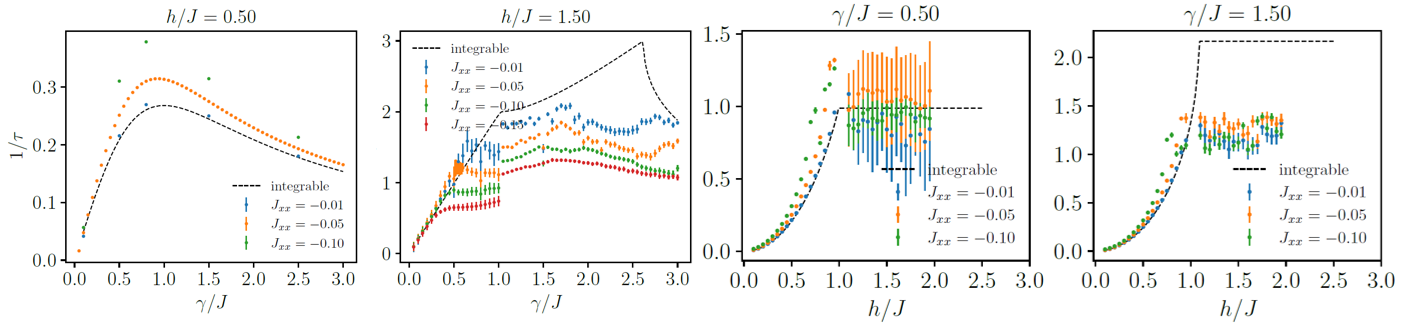}
\caption{Decay constants for $XX$-interacting Ising chain, obtained by fitting either exponential or power-law-dressed exponential. 
We force a power-law-dressed fit in $h>1$ and $\gamma<1$ region, resulting in abrupt jumps near this point where the fitting 
procedure changes.}
\label{fig_final_plots_novel_fit_H2} 
\end{figure}

To confirm robustness of the results for various integrability-breaking
interactions, in this section we consider the Hamiltonian 
\begin{equation}
H_{xx}=-J\sum_{j}\sigma_{j}^{z}\sigma_{j+1}^{z}-h\sum_{j}\sigma_{j}^{x}+J_{xx}\sum_{j}\sigma_{j}^{x}\sigma_{j+1}^{x}\label{eq_H2}
\end{equation}
which also breaks integrability, but without the self-dual interaction
that restricts the phase transition to $h=J$. Some data for this
are shown in Figure \ref{fig_final_plots_novel_fit_H2}. The general behavior is similar to that of the self-dual interaction.
Sharp kinks in $\tau^{-1}$ persist, but no longer at the same apparent critical points due to the lack of self-duality. This also
causes issues with our fitting procedure, as we manually enforce a power-law-dressed exponential fit for $h>1$ and $\gamma < 1$. 
Since these are apparently no longer the locations of the boundary critical points, a better fitting procedure should be 
adopted in future work.

\bibliography{references.bib}

\end{document}


\title{Zeno physics of the Ising chain with symmetry-breaking boundary dephasing}

\author{Umar Javed$^{1}$, Riccardo J. Valencia-Tortora$^{2}$, Jamir Marino$^{2}$, Vadim Oganesyan$^{3,4}$, Michael Kolodrubetz$^1$}
\affiliation{$^1$Department of Physics, The University of Texas at Dallas, Richardson, Texas 75080, USA}
\affiliation{$^2$Institut f\"{u}r Physik, Johannes Gutenberg-Universit\"{a}t Mainz, D-55099 Mainz, Germany}
\affiliation{$^3$Physics program and Initiative for the Theoretical Sciences, The Graduate Center, CUNY, New York, NY 10016, USA}
\affiliation{$^4$Department of Physics and Astronomy, College of Staten Island, CUNY, Staten Island, NY 10314, USA}

\begin{abstract}

In few-qubit systems, the quantum Zeno effect occurs when measurement occurs sufficiently frequently that the spins are unable to relax between measurements. This can compete with Hamiltonian terms, resulting in interesting relaxation processes which can be captured by two-time correlation functions evolving under the Lindblad equation. However, boundary dynamics are known to be significantly affected by bulk phase transitions, as in the canonical transverse field Ising chain. In this work, we study the effect of bulk properties on the boundary quantum Zeno effect. We find that sharp singularities occur in the boundary relaxation dynamics, which can be tied to the emergence or destruction of edge modes. We analytically describe these edge modes and their relation to the spin dynamics. We provide numerical evidence that the dynamical singularities are stable in the presence of integrability-breaking interactions.

\end{abstract}

\maketitle

\begin{figure}[b]
    \centering
    \includegraphics[width=0.7\columnwidth]{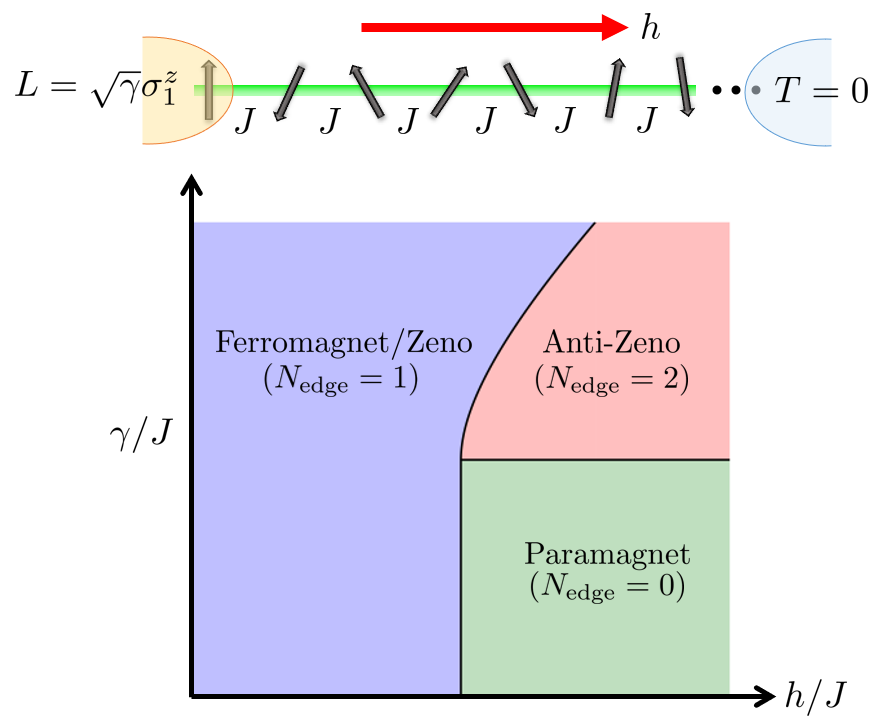}
    \caption{Illustration of Ising model with symmetry-breaking boundary noise. Far from the boundary, the Ising model is in its ground state ($T=0$). Boundary spin dynamics show singularities which can be connected to edge modes, resulting in a nontrivial boundary phase diagram.}
       \label{fig:model}
\end{figure}

Understanding the nature of correlated phases of matter in many-body systems beyond the perturbative regime is challenging, even more so if the correlated system is out of equilibrium. Yet nonequilibrium quantum physics is precisely the regime that is increasingly experimentally accessible in systems ranging from cold atoms and ions to superconducting circuits to (noisy) quantum computers \cite{Langen2015,Bruzewicz2019,Kjaergaard2020}. Therefore, characterizing nonequilibrium quantum systems and their phases of matter is one of the most important questions in many-body physics today.

Given the lack of solvable non-equilibrium models, it is crucial to develop an in-depth understanding of the few models in which analytical methods exist. One important class of such systems are quantum boundary/impurity models. Examples where these can be treated beyond the perturbative limit include the Anderson impurity~\cite{nozieres1969singularities,anderson1967infrared,silva2008statistics,cetina2016ultrafast,knap2012time,tonielli2019orthogonality},  transport in one-dimensional junctions with strong electron-electron interactions ~\cite{kane1992transport,kane1995impurity}, and the generation and growth of entanglement among magnetic impurities and their surrounding environments in the Kondo effect ~\cite{latta2011quantum, cox1998exotic, andrei1995}. A key tool for addressing many of these example systems is boundary conformal field theory (CFT), which uses generalized scale invariance near quantum critical points to build a significant analytical toolkit \cite{PhysRevB.96.241113,Vasseur,Francica,Prev_2,Kun,campostrini2014finite,Campostrini}.

However, even boundary CFT is not equipped to handle an increasingly important class of systems: open quantum systems. Motivated by developments in quantum simulation and computation, it is increasingly important to consider quantum systems which are driven far-from-equilibrium yet, inevitably, remain coupled to their environments. Despite hindering some goals of quantum information theory, dissipation has been found to be useful in other ways such as aiding in state preparation \cite{Budich2015,Reiter2016,Harrington2022} and enhancing the phase structure of quantum matter \cite{Diehl2008,sieberer2016keldysh,Rakovszky2023}. Nonequilibrium open quantum systems are even harder to treat in the strongly correlated regime, necessitating the study of models where concrete predictions can be made.

In this paper, we show that such a model is the transverse field Ising chain with symmetry-breaking boundary dephasing. Despite symmetry breaking at the boundary, it remains integrable, with certain correlation functions expressible in the language of free Majorana fermions. Edge modes emerge which are naturally described in terms of the Majoranas, giving rise to an interesting (boundary) phase diagram. Furthermore, we show that these edge modes are directly connected to boundary spin dynamics, such that the appearance or disappearance of edge modes coincides with sharp changes in the dynamics. We argue that these singularities are robust to integrability-breaking interactions in the appropriate scaling limit and support the argument with matrix product state-based numerics. Finally, we argue that this is a fundamentally nonequilibrium phenomenon that is hidden from conventional equilibrium observables. We show this explicitly for the steady state energy current from the bath and boundary magnetic susceptibility.

\section{Model}

We consider the one-dimensional Ising model with open boundary conditions and strong dephasing that breaks Ising symmetry at the boundary. Specifically, we assume dissipative time evolution of the Lindblad form, 
\begin{equation}
    \frac{d \rho}{dt}=\mathcal{L}[\rho]=-i[H,\rho]+ L \rho L ^{\dagger} -\frac{1}{2}\{L^{\dagger} L ,\rho\},
\end{equation}
with Hamiltonian 
\begin{equation}
    H=-J\sum_{n=1} ^{L-1} \sigma_{n} ^z \sigma_{n+1} ^z - h\sum_{n=1} ^{L} \sigma_n ^x
\end{equation}
and a single Lindblad operator
\begin{equation}
    L=\sqrt{\gamma} \sigma_1^z.
\end{equation}
We use the convention $J=\hbar=1$ throughout.

We are interested in the dynamics of the boundary spin, $\sigma^z_1$, in the presence of this boundary dissipation. One motivation for this is the quantum Zeno effect, which for $J=0$ (single spin) says that the spin dynamics get frozen in the limit $\gamma \gg h$ due to repeated measurements by the environment. By adding interactions $J>0$, we wish to understand the effect of many-body physics -- including quantum phase transitions -- on Zeno physics. Furthermore, the phase transition in the Ising model is a canonical example of a conformal field theory (CFT). By coupling this system to a noisy, relevant boundary perturbation, there is hope to draw a connection between many-body Zeno physics and boundary CFT. Similar ideas have been explored in recent works \cite{berdanier2019universal,froml2019fluctuation,dolgirev2020non}, but with notable differences in the model or the dynamical signature \ric{\textbf{[I would elaborate more on the differences to highlight our contribution and better frame our work]}}.

In the absence of static symmetry-breaking boundary field, the spin itself is unpolarized due to symmetry. Therefore, the quantity of interest is the two-time correlation function, which is related to the boundary magnetic susceptibility. Specifically, we seek to find the two-time correlation function in the time-evolved density matrix $\rho(t)$, which is given by using the regression theorem \cite{howard}:
\begin{equation}
    C(t,\Delta t)=\langle \sigma_1 ^z (t+\Delta t) \sigma_1 ^ z (t) \rangle =\mathrm{Tr}\left\{ \left[ e^{\mathcal{L^\dagger} \Delta t} \sigma_1^z \right] \sigma_1^z \rho (t) \right\},
\end{equation}
where the adjoint Liouvillian $\mathcal{L}^\dagger$ generates Heisenberg operator evolution. The initial state is $\rho(0) = |\psi_\mathrm{gs}\rangle \langle \psi_\mathrm{gs} |$, the ground state of the unperturbed Ising chain ($\gamma=0$). Then, at $t=0$, a finite value of $\gamma$ is quenched on. Once we introduce this noise at the edge, we expect quasiparticle excitations to travel ballistically to the other edge of the chain in time $t\approx L/v$. We are interested in the non-equilibrium steady state (NESS) that forms near the boundary for $t + \Delta t < L / v$. In order to equilibriate to the local NESS, $t$ must be chosen to be sufficiently large as well; we use $t=40$ throughout our data. Note that this is equivalent to the true NESS that forms for a large, zero temperature bath placed sufficiently far from the perturbed boundary (see Figure \ref{fig:model}).

Numerically calculating the two-time autocorrelation function remains challenging in general, but for this particular model it can be efficiently obtained by rewriting in terms of Majorana fermions. We start by a conventional Jordan-Wigner transform on the spin degrees of freedom \cite{sachdev_2011}, 
\begin{align}
    \sigma_j^x =i \eta_{2j-1} \eta_{2j},\ \sigma_j^z = \left(\prod_{n=1} ^{j-1} i \eta_{2n-1} \eta_{2n}\right) \eta_{2j-1}.
\end{align}
The boundary spin is then a single Majorana operator, $\sigma_1^z=\eta_{1}$, and the Hamiltonian can be written \ric{as}
\begin{equation}
    H=- i h \sum_{j=1} ^{L} \eta_{2j-1} \eta_{2j} - iJ\sum_{j=1}^{L-1}\eta_{2j} \eta_{2j+1}.
\end{equation}
The first step is to evolve the density matrix to $\rho(t)$ and calculate the equal-time correlation function $C(t,0)=\mathrm{Tr}\left[  \sigma_1^z \sigma_1^z \rho (t) \right] = \mathrm{Tr}\left[  \eta_{1}^2 \rho (t) \right]$. Since $\eta_{1}^2=1$, this is always equal to $1$. Crucially, we can also calculate the full Majorana correlation matrix $\mathcal{C}_{ij}(t,0) = \mathrm{Tr}\left[  \eta_{i} \eta_j \rho (t) \right]$ such that $C=\mathcal{C}_{11}$. Two-time correlation functions are then able to be calculated because the adjoint Liouvillian conserves Majorana number:
\begin{equation}
\mathcal{L}^\dagger \left[ \eta_{i} \right] =i[H,\eta_{i}]+\gamma(\eta_{1}\eta_{i}\eta_{1}-\eta_{i}) = \sum_{j\ric{=1}}^{\ric{2L}} M_{ij} \eta_j.
\end{equation} 
The matrix $M$ has the form
\begin{equation}
   M=\left(\begin{array}{cccc}
0 & -2h & 0 & 0\\
2h & -2\gamma & -2J & 0\\
0 & 2J & -2\gamma & -2h\\
0 & 0 & 2h & \ddots
\end{array}\right).
\end{equation}
One can then readily show that the 2-time correlation matrix evolves as
\begin{equation}
    \frac{\partial \mathcal{C}(t,\Delta t)}{\partial \Delta t} = M \mathcal{C}(t,\Delta t)
\end{equation}
with equal-time correlations, $\mathcal{C}(t,0)$, as an initial condition. Clearly the eigenmodes of $M$ play a crucial role in understanding the dynamics of the boundary spin; we therefore refer to this as the single-Majorana evolution matrix. More details for how these equations are solved numerically may be found in the supplement \ref{sec:sup_numerics}.

We note briefly that an alternative route to solving the dissipative dynamics exists by vectorizing the density matrix and treating the Lindblad evolution via third quantization. This doubles the effective Hilbert space, but manifestly makes the problem integrable because the boundary dephasing maps to a non-Hermitian Ising term connecting the bra and ket Hilbert spaces. A similar technique was used in recent papers \cite{Shibata2020,Zheng2023}, which studied the related problem of dephasing connected to both ends of a finite system and found a similar phase diagram of the edge modes. While formally identical to our method, third quantization makes it more challenging to connect spectral features and eigenmodes of the Liouvillian to physical observables. Therefore, we are able to more clearly study physically observable properties in this work and make stronger statements about broader relevance and universality. More details on the third quantization formalism and its connection to the single-particle evolution matrix $M$ can be found in the supplementary information in sections \ref{sec:supp_third_quantization}  and \ref{sec:supp_connecting_second_and_third_quantization}.

\section{Results}

\begin{figure}
    \centering
    \includegraphics[width=\columnwidth]{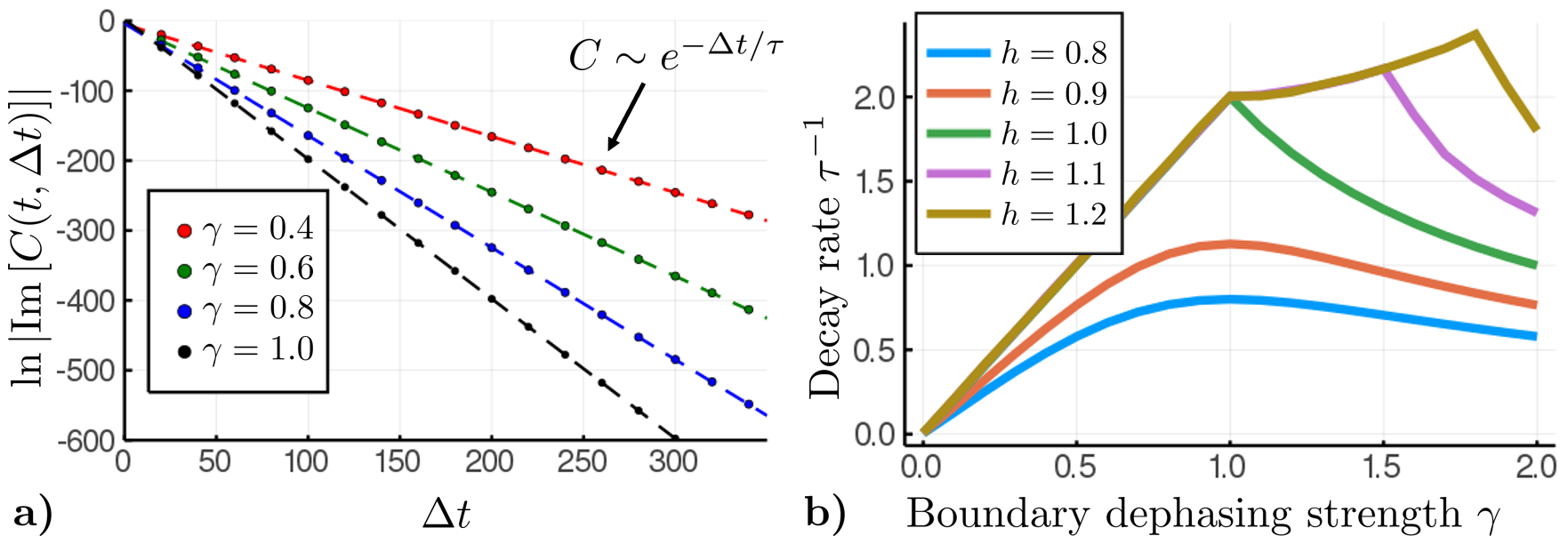}
    \caption{(a) Numerical results for two-time correlation function of dissipative edge spin as a function of boundary dissipation strength ($\gamma$) for $L=1000$, $J=h=1$, and $t=40$. (b) Plot of decay rate $\tau^{-1}$ as a function of $\gamma$ for different values of $h$. }
       \label{fig:correlation_function}
\end{figure}

Using the free fermion\ric{s} form, we numerically solve for the 2-time correlations of the boundary spin. For all parameters chosen, the (complex-valued) autocorrelation function is found to decay exponentially \ric{at late times $\Delta t$ as} $C\ric{\sim}C_0 \exp[-\Delta t/\tau]$, where $\tau$ is the relaxation time scale the characterizes the Zeno effect. 
We fit the numerical data for large $\Delta t$ to extract a value of $\tau$, as shown in Figure \ref{fig:correlation_function} a.

Plotting the relaxation rate $\tau^{-1}$ as a function of $\gamma$ and $h$ (Figure~\ref{fig:correlation_function}b), we find the striking result that, for $h \geq 1$, sharp singularities occur in this decay constant \ric{in a fashion \cancel{which are}} reminiscent of equilibrium observables in a conventional first order phase transition. These singularities separate the parameter space into three distinct phases (Figure \ref{fig:model}, bottom), which we label paramagnet (PM), ferromagnet (FM)/Zeno, and anti-Zeno. The PM and FM smoothly connect to the respective ground state phases at $\gamma=0$ with a transition that extends vertically from $h=J$. The anti-Zeno phase only appears at $\gamma > J$ and, as we will see, cannot be thought of as smoothly connected to the ground state Ising physics.

\begin{figure*}
    \centering
    \includegraphics[width=0.85\textwidth]{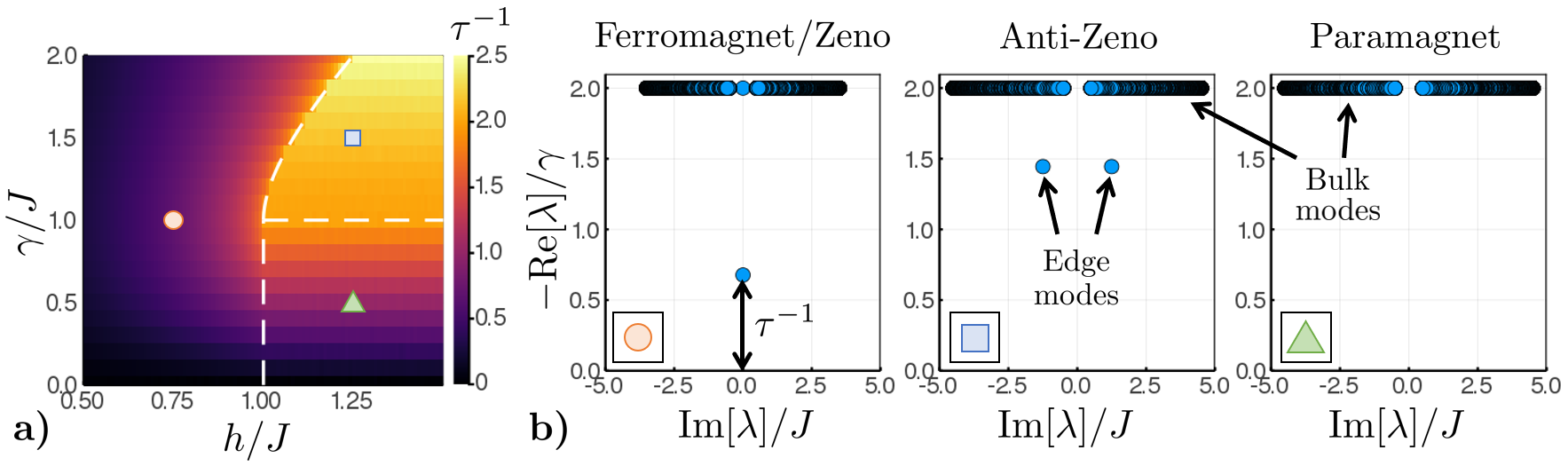}
    \caption{(a) Phase diagram obtained from singularities of the relaxation rate $\tau^{-1}$. (b) Eigenvalues $\lambda$ of the Majorana evolution matrix $M$ plotted for three different points in the phase diagram (left panel), showing number and structure of edge modes. The decay rate is set by the mode with the smallest magnitude real component.}
    \label{fig:spectrum}
\end{figure*}

In order to understand the origin of these singularities, an interesting analogy can be drawn to our previous work \cite{javed2023counting}, which examined the Ising chain with a static symmetry-breaking boundary field. There, we observed a connection between the two-time correlation function of the edge spin and emergent edge modes in the Majorana problem. While this setup is qualitatively different, we are inspired to ask a similar question of the effective non-Hermitian matrix $M$ which determines time evolution of the Majorana correlations. We begin by solving the eigenmodes of $M$ numerically, results of which are shown in Figure \ref{fig:spectrum}b. There are clearly modes that appear outside of the bulk, which we confirm are edge modes localized near the dissipative site. Furthermore, we confirm that the change in edge mode counting is directly tied to the singularities in $\tau$.

In fact, we can go one step further and solve for the edge modes of $M$ as well as their phase transitions analytically. We consider the following (unnormalized) ansatz for the edge mode
\begin{equation}
\eta_{\text{edge}}=\sum_{j=1}^{N}\left(r^{j-1}\eta_{2j-1}+Ar^{j-1}\eta_{2j}\right),
\end{equation}
where $\Big| r=e^{-1/\xi + i\varphi} \Big| < 1$  describes the decay with length scale $\xi > 0$.  We are interested in the ``eigenenergy'' $\lambda$  such that $M\eta_\mathrm{edge}=\lambda \eta_\mathrm{edge}$.  This equation is solved in the section~\ref{sec:supp_edge_modes_M} \ric{from which we observe that the number of physically acceptable edge modes ($|r| < 1$) changes in correspondence with \cancel{giving}} the phase boundary
\begin{equation}
    h_c = \begin{cases}
        J & \mathrm{for}~\gamma < J \\
        \frac{\gamma^{2}+J^2}{2\gamma} & \mathrm{for}~\gamma > J
    \end{cases}
\end{equation}

Moreover, this analytical solution provides a complete description of the edge modes and their phase transitions, allowing us to re-interpret the observed dynamical signatures physically. For example, we readily see that the PM and FM edge modes track continuously to the Majorana zero modes of the Kitaev chain as $\gamma \to 0$.  As $h$ approaches $J$ from the FM side, the edge mode gap approaches the bulk dissipation rate of $2\gamma$, such that the edge mode merges into the bulk simultaneously with the bulk gap closing at $h=J$. The phase transition looks similar to the conventional ground state Ising transition, with diverging edge correlation length $\xi \to \infty$ and a dissipative gap $\Delta \sim |h-J|$  suggesting that $\nu=z=1$ as in the ground state.

Another relatively simple limit of the model is $\gamma \gg J$, for which the physics of the single boundary spin becomes dominant. In this case, the limits $\gamma \ll h$ and $\gamma \gg h$  corresponds to the anti-Zeno and Zeno regimes, respectively. While there is not a conventional phase transition in the steady state of the single-spin Zeno model, there is an exceptional point (i.e. under- to over-damped) transitions at $\gamma_c=2h$.  This is precisely the asymptote that we find for our phase boundary when  $\gamma,h \gg J$.  However, all the way down to $\gamma=J$ there is an exceptional point transition in the spectrum of $M$, allowing us to argue that one has Zeno and anti-Zeno ``phases.'' 

The final boundary transition happens at $\gamma=J$ for $h>J$, between the PM and anti-Zeno phases. At this transition, two non-Hermitian bound states emerge out of the PM continuum at finite momentum. The critical behavior appears Ising-like, with imaginary gap that opens linearly with the tuning parameter $\gamma$ ($\nu=z=1$). Furthermore, the entire transition line has diverging correlation length $\xi \to \infty$ (i.e., $|r|\to 1$) but non-zero momentum $\varphi \neq 0$ except at the tricritical point $h=J=\gamma$. While this transition occurs at finite dephasing strength and away from bulk criticality, this diverging length scale suggests that a appropriately defined field theory with gapped bulk may be possible to construct, for which the transition may be universal.

\section{Interacting System}

\begin{figure*}
\centering
\includegraphics[width=0.8\textwidth]{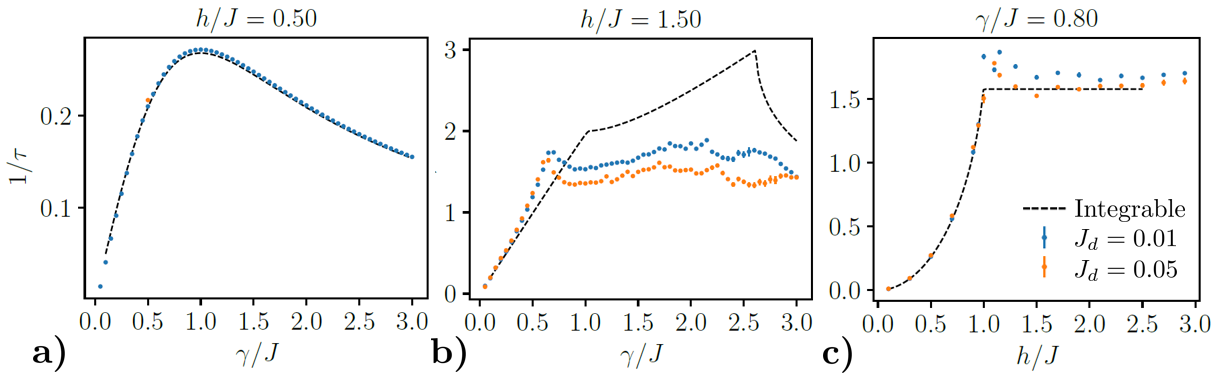}
\caption{Relaxation rate extracted from TEBD simulations of the Ising chain with self-dual interactions. The results are given by fitting either a pure exponential $e^{-\Delta t/\tau}$ or one dressed by a power law $(\Delta t)^{-3/2} e^{-\Delta t/\tau}$ in the $h>1$ and $\gamma<1$ region.  The dashed lines in the plots $1/\tau$ are the results from the integrable case. For small $\gamma$, oscillations are still clearly present in $C(t,\Delta t)$ for the small $\Delta t$ available, making the fitting procedures less accurate.}
\label{fig_final_plots_novel_fit}
\end{figure*}

In order to probe universality of this phase diagram, we now study the robustness of this edge physics in the presence of integrability-breaking interactions. We will do so by modifying the Hamiltonian to 
\begin{equation}
\label{eq_H3}
\begin{split}
H_\mathrm{int} =& -J \sum_{j} \sigma^z_j \sigma^z_{j+1} - h \sum_j \sigma^x_j +\\
&- J_d\sum_{j} \left(\sigma^x_j \sigma^x_{j+1}+\sigma^z_j \sigma^z_{j+2}\right).
\end{split}
\end{equation}
The form of the interaction is chosen such that it is invariant under the Kramers-Wannier (spin/bond) duality, meaning that if a ground state phase transition exists in the bulk, it must occur at the self-dual point $h=J$. 

Since free fermion methods do not work for this interacting model, we must resort to more challenging numerics. Specifically, we solve the dynamics via a variant of time-evolving block decimation (TEBD) in which the density matrix is treated in third quantization and unfolded such that boundary dissipation becomes a non-Hermitian perturbation at the center of the spin chain (see supplement \ref{sec:supp_third_quantization}). We measure the autocorrelation functions for various $J_d > 0$ (a ferromagnetic perturbation) after first evolving to time $t=1/J$. We consider system size $L=30$, which is sufficient to approximate $L=\infty$ for the timescales we are able to simulate, since no  appreciable finite-size effects occur before $t \sim L/J$ .

We tested our methods by first applying them to the integrable case $J_d = 0$ (see \ref{sec:tebd_non_interacting}) to confirm that it reproduced the known analytical and numerical results from free fermions. One issue that arises in the TEBD numerics is the relatively small $\Delta t \sim 15$ that can be accessed numerically. Specifically, within the PM phase a simple exponential does not fit the data well. The origin of this mismatch comes from the absence of edge modes, meaning that relaxation in the PM comes from the bulk continuum. However, after extracting the exponential prefactor, the remaining power law decay of the boundary spin correlations is strikingly similar to that of the Ising chain with static boundary field \cite{javed2023counting}. Calculating this decay via saddle point approximation gives a $t^{-3/2}$ multiplying the exponential, which is found to give a much cleaner fit and therefore reproduce $\tau^{-1}$ more accurately.

We then apply the same fitting procedure to the values of $C(t,\Delta t)$ obtained in the presence of integrability-breaking interactions. The results (in Figure~\ref{fig_final_plots_novel_fit}) are far from clear, but do show strong signatures of the same phase transitions at both the PM/anti-Zeno and FM/PM transitions. This suggests that the transitions do indeed survive on this time scale despite the presence of integrability breaking. Furthermore, while the data does not resolve a singularity between the FM/Zeno and anti-Zeno phase \ric{due to the accessible short times}, that limit is expected to be the most robust to interactions because it occurs on finite length scale $\xi$, which becomes particularly small in the limit $h,\gamma\gg J$. We therefore speculate that all the transitions survive interactions on this time scale, though detecting them may be challenging.

At late time $t+\Delta t \gg L / v$, one expects the system to heat up. With just boundary dephasing, the eventual steady state would correspond to infinite temperature. However, there are a few caveats which open the possibility that some of these phenomena will be infinitely long-lived. First, if the chain is semi-infinite, then time $L/v$ can never be reach. This is quite similar to the situation shown in Figure~\ref{fig:model} in the sense that the semi-infinite chain can be thought of as an unbounded zero temperature heat sink. If the mean free path $\ell_\mathrm{mfp}$ is much longer than the edge mode localization length $\xi$, then one would expect these excitations to be carried out of the system by the zero temperature bath before they can backscatter and thermalize the edge mode. The precision of this statement, including whether it turns transitions into crossovers, remains unclear; it will be an important topic for future investigation.

\section{Equilibrium observables}

Having identified the emergence of edge modes that are tied to the dynamical singularities, we may expect that signatures of these edge modes will also show up in other observables signifying properties of the nonequilibrium steady state. In the supplementary material, we study magnetic susceptibility and energy current in the NESS (sections \ref{sec:supp_magn_susc} and \ref{sec:supp_energy_current}). We find no evidence for singularities of either quantity. This is tied to the evolution of the equal-time correlation matrix $\mathcal{C}(t,\Delta t=0)$, which is governed by a matrix similar to $M$ but with exponential decay of the edge rather than bulk. This reduces the impact of the edge modes in quasi-equilibrium observables and, as we argue in the supplement, appears to prevent singularities in these quantities. Therefore, we conclude that the edge transitions are a dynamical phenomenon which are not captured by  conventional equilibrium properties.

\section{Conclusion}

We have investigated the transverse field Ising chain in the presence of boundary dissipation. We calculated the decay rate of two-time correlations of the boundary spin and showed that sharp singularities occur in between the Zeno and anti-Zeno regimes. The corresponding phase diagram divides into three distinct phases whose boundary relaxation dynamics are directly tied to analytically solvable edge modes in the related Majorana problem. This phase diagram appears robust to interactions in TEBD simulations, consistent with our prediction that the dynamical phenomena will be universal within one-dimensional models with Ising symmetry.

Our results have potentially far-reaching implications if they can be generalized to dissipative versions of other well-studied models, most notably boundary CFTs with dephasing that couples to a relevant boundary perturbation. Applying the same renormalization group machinery near the critical point of a more complex CFT suggests that boundary dephasing can be a relevant perturbation, as opposed to the marginally irrelevant case that we found here \ric{\textbf{[I am a bit confused, in Page 2 first column, we say it is a relevant boundary perturbation. Am I missing something?]}}. In particular, Potts models with $\mathds{Z}_q$  symmetry has increasingly relevant boundary perturbations for $q\geq 3$ \cite{Cardy1984_bcft_dims}. This likely implies the creation of a new boundary phase near the critical point; our work to find this behavior remains on-going. Similar dissipative boundary terms can be studied for a wide class of models, including experimentally relevant situations such a dissipative Kondo problem which is known to be equivalent to a boundary CFT \cite{Affleck1995}. Finally, we note that dissipative impurity problems are naturally created in a variety of experimental settings where their dynamics remain challenging to solve. One important example of this is resonant inelastic x-ray scattering (RIXS), in which a localized ``core hole'' impurity modifies the dynamics of the itinerant electrons around it. The problem naturally involves relaxation dynamics as electrons attempt to fill this hole. Therefore, we hope that some of the techniques that we have developed here for an analytically tractable dissipative Ising model can be generalized to this more challenging, but experimentally relevant, setting.

\onecolumngrid

\section*{Supplementary information}

In this supplement, we provide additional information and data on Majorana time evolution, analytical solutiosn for the edge states, third-quantized treatments of the dynamics, equilibrium observables, and computational methods for the integrability-breaking model.

\mkaddcomment{TO DO LIST:
\begin{enumerate}
    \item Fix missing citations (search for ? and hl).
    \item Clean up edge state calculations
    \item Make sure $M$  and $\mathcal H$  matrices are correct
    \item Add argument for PM/Anti-Zeno phase transition
    \item Change terminology throughout to label phases ``FM/Zeno'', ``PM'', and ``Anti-Zeno''
    \item Call the Majoranas $\eta_1$, $\eta_2$, etc. Don't use $A$ and $B$.
\end{enumerate}
}

\section{Details regarding second-quantized Majorana time evolution }

Throughout the main text, we primarily solve Majorana correlation functions using their direct Heisenberg time evolution. In this section, we provide more details on how this is done numerically. We also solve for the edge states analytically in a semi-infinite geometry.

\subsection{Numerics for 1- and 2-time correlation functions \label{sec:sup_numerics}}

The time evolution of our system is given by the
Lindblad equation

\begin{equation}
\frac{d\rho}{dt}=\mathcal{L}[\rho]=-i[H,\rho]+L\rho L^{\dagger}-\frac{1}{2}\{L^{\dagger}L,\rho\}
\end{equation}
where $\mathcal{L}$ is the Liouvillian super-operator and $L$ is a Lindblad operator which has the form 
\begin{equation}
L=\sqrt{\gamma}\sigma_{1}^{z}
\end{equation}
Using the Jordan-Wigner transformation, we can write our Hamiltonian
and Lindblad operator in terms of Majorana fermions as: 
\begin{align}
H & =-iJ\sum_{n=1}^{L-1}\eta_{2n}\eta_{2n+1}-ih\sum_{n=1}^{L}\eta_{2n-1}\eta_{2n}\equiv i\eta^{T}A\eta\\
L & =\sqrt{\gamma}\eta_{1}
\end{align}
where $A$ is an $L\times L$ real antisymmetric matrix and $\eta=(\eta_{1},\eta_{2},\eta_{3},\ldots)^{T}$
is a column vector of Majorana operators.

We are interested in calculating a 2-time correlation function of
the Majoranas starting from initial state $\rho(t=0)$, which is given
by the regression theorem \cite{howard} as 
\begin{equation}
\mathcal{C}_{ij}(t,\Delta t)=\mathrm{Tr}\left[\eta_{i}e^{\mathcal{L}\Delta t}\left(\eta_{j}\rho(t)\right)\right]\label{eq:C_t_Delta_t}
\end{equation}
First, let's calculate the time evolution of the equal-time correlation
function, $\Delta t=0$. Note that the diagonal elements are always
equal to 
\begin{equation}
\mathcal{C}_{jj}\left(t,0\right)=\mathrm{Tr}\left[\eta_{j}^{2} \rho(t)\right]=1.
\end{equation}
The time derivative of the off-diagonal terms ($i\neq j$) is 
\begin{align}
\frac{d\mathcal{C}_{ij}}{dt} & =\text{Tr}\left[\eta_{i}\eta_{j}\frac{d\rho}{dt}\right]\\
 & =\text{Tr}\left[\eta_{i}\eta_{j}\left(-i[H,\rho]+L\rho L^{\dagger}-\rho\right)\right]\\
 & =\text{Tr}\left[\eta_{i}\eta_{j}\left(\sum_{kl}A_{kl}[\eta_{k}\eta_{l},\rho]+\gamma\eta_{1}\rho\eta_{1}-\gamma\rho\right)\right]
\end{align}
Despite appearing to involve 4-point correlation functions of the
Majoranas, we can simplify the above equation to only involve 2-point
correlations by using the anticommutation relations for our Majorana
operators, $\{\eta_{i},\eta_{j}\}=2\delta_{ij}$, as well as cyclicity
of the trace and antisymmetry of $A$ and $\mathcal{C}$: 
\begin{align}
\sum_{kl}A_{kl}\text{Tr}\left(\eta_{i}\eta_{j}[\eta_{k}\eta_{l},\rho]\right) & =\sum_{kl}A_{kl}\text{Tr}\left(\left[\eta_{i}\eta_{j},\eta_{k}\eta_{l}\right]\rho\right)\\
 & =\sum_{kl}A_{kl}\left\langle\eta_{i}\eta_{j}\eta_{k}\eta_{l}-\eta_{k}\eta_{l}\eta_{i}\eta_{j}\right\rangle\\
 & =\sum_{kl}A_{kl}\left\langle \eta_{i}\eta_{j}\eta_{k}\eta_{l}-2\delta_{il}\eta_{k}\eta_{j}+\eta_{k}\eta_{i}\eta_{l}\eta_{j}\right\rangle \\
 & =\sum_{kl}A_{kl}\left\langle \eta_{i}\eta_{j}\eta_{k}\eta_{l}-2\delta_{il}\eta_{k}\eta_{j}+2\delta_{ik}\eta_{l}\eta_{j}-\eta_{i}\eta_{k}\eta_{l}\eta_{j}\right\rangle \\
 & =\sum_{kl}A_{kl}\left\langle \eta_{i}\eta_{j}\eta_{k}\eta_{l}-2\delta_{il}\eta_{k}\eta_{j}+2\delta_{ik}\eta_{l}\eta_{j}-2\delta_{jl}\eta_{i}\eta_{k}+\eta_{i}\eta_{k}\eta_{j}\eta_{l}\right\rangle \\
 & =\sum_{kl}A_{kl}\left\langle -2\delta_{il}\eta_{k}\eta_{j}+2\delta_{ik}\eta_{l}\eta_{j}-2\delta_{jl}\eta_{i}\eta_{k}+2\delta_{jk}\eta_{i}\eta_{l}\right\rangle \\
 & =2\sum_{kl}\left[\delta_{il}\mathcal{C}_{jk}A_{kl}+\delta_{ik}A_{kl}\mathcal{C}_{lj}-\delta_{jl}\textbf{C}_{ik}A_{kl}-\delta_{jk}A_{kl}\textbf{C}_{li}\right]\\
 & =4\left(A\textbf{C}-\textbf{C}A\right)_{ij}\\
\text{Tr}\left[\eta_{i}\eta_{j}\eta_{1}\rho\eta_{1}\right]-\text{Tr}\left[\eta_{i}\eta_{j}\rho\right] & =\text{Tr}\left[\eta_{1}\eta_{i}\left(2\delta_{1j}-\eta_{1}\eta_{j}\right)\rho\right]-\textbf{C}_{ij}\\
 & =2\delta_{1j}\mathcal{C}_{1i}-\text{Tr}\left[\eta_{1}\left(2\delta_{1i}-\eta_{1}\eta_{i}\right)\eta_{j}\rho\right]-\mathcal{C}_{ij}\\
 & =2\delta_{1j}\mathcal{C}_{1i}-2\delta_{1i}\mathcal{C}_{1j}\\
 & =-2\left(\delta_{1j}+\delta_{1i}\right)\mathcal{C}_{ij}=-2\left\{ \Delta_{1},\mathcal{C}\right\} _{ij}\\
\implies\frac{d\mathcal{C}}{dt} & =4\left[A,\mathcal{C}\right]-2\gamma\left\{ \Delta_{1},\mathcal{C}\right\} +4\gamma\Delta_{1}\mathcal{C}\Delta_{1}\label{eq:final_eqn_dC_dt}
\end{align}
where 
\begin{equation}
\Delta_{1}=\left(\begin{array}{cccc}
1 & & & \\
 & 0 & & \\
 &  & 0 & \\
 &  &  & \ddots
\end{array}\right)
\end{equation}
projects onto the first row/column. We also used that 
\begin{equation}
CA=\left(-C^{T}\right)\left(-A^{T}\right)=\left(AC\right)^{T}
\end{equation}
The above calculation is a specific example of the general fact that
our Liouvillian has the form of free Majorana time evolution, and
therefore can be efficiently simulated. Since the right hand side
of Eq.~\ref{eq:final_eqn_dC_dt} is linear in $\mathcal{C}$, we can also vectorize
$\mathcal{C}$ and write this as $\dot{\mathcal{C}}=\mathcal{M}\mathcal{C}$, where $\mathcal{M}$
is a supermatrix. Then the solution is $\mathcal{C}(t)=e^{\mathcal{M}t}\mathcal{C}(0)$.

Having calculated $C(t,0)$, the 2-time correlation function is relatively
easy, involving similar methods. First note that Eq.~\ref{eq:C_t_Delta_t}
can be thought of as time evolution of $\rho\eta_{j}$ under $\mathcal{L}$
by $\Delta t$, where initial conditions are set by $C(t,0)$. A slightly
simpler expression is obtained by moving the time evolution over to
$\eta_{i}$ using the adjoint of the Liouvillian, which satisfies
the property that 
\begin{equation}
\text{Tr}\left[A\cdot\mathcal{L}(B)\right]=\text{Tr}\left[\mathcal{L}^{\dagger}(A)\cdot B\right].
\end{equation}
For Hermitian Lindblad operator, as we have here, $\mathcal{L}^{\dagger}$ is identical to $\mathcal{L}$ except with $H\to -H$.
Therefore,
\begin{equation}
\mathcal{C}_{ij}(t,\Delta t)=\mathrm{Tr}\left[e^{\mathcal{L}^\dagger\Delta t}\left(\eta_{i}\right)\eta_{j}\rho(t)\right]\implies\frac{\partial \mathcal{C}_{ij}(t,\Delta t)}{\partial\Delta t}=\mathrm{Tr}\left[\mathcal{L}^\dagger\left(\eta_{i}\right)\eta_{j}\rho(t)\right].
\end{equation}
 Then, via similar machinery as earlier, ***here***
\begin{align}
\mathcal{L}^\dagger\eta_{i} & =i[H,\eta_{i}]+\gamma(\eta_{1}\eta_{i}\eta_{1}-\eta_{i})\\
 & =-\sum_{kl}A_{kl}\left[\eta_{k}\eta_{l}\eta_{i}-\eta_{i}\eta_{k}\eta_{l}\right]+\gamma(\eta_{1}\eta_{i}\eta_{1}-\eta_{i})\\
 & =-\sum_{kl}A_{kl}\left[\eta_{k}\left(2\delta_{il}-\eta_{i}\eta_{l}\right)-\eta_{i}\eta_{k}\eta_{l}\right]+\gamma\left(2\delta_{i1}-\eta_{i}\eta_{1}\right)\eta_{1}-\gamma\eta_{i}\\
 & =-\sum_{kl}A_{kl}\left[2\delta_{il}\eta_{k}-\left(2\delta_{ik}-\eta_{i}\eta_{k}\right)\eta_{l}-\eta_{i}\eta_{k}\eta_{l}\right]+\gamma\left(2\delta_{i1}-\eta_{i}\eta_{1}\right)\eta_{1}-\gamma\eta_{i}\\
 & =4\sum_{k}A_{ik}\eta_{k}+2\gamma\eta_{i}\left(\delta_{i1}-1\right)\\
\frac{\partial C(t,\Delta t)}{\partial\Delta t} & =4AC+2\gamma\left(\Delta_{1}-\mathds1\right)C\\
\implies C(t,\Delta t) & =\exp\left[4A+2\gamma\left(\Delta_{1}-\mathds1\right)\right]C(t,0)
\end{align}

\hl{[Is the remaining piece of this repetitive?]}

We can write the two time correlation in the following form (shown in supplemental material)
\begin{equation}
    C(t,\Delta t)  =\exp\left[M \Delta t\right]C(t,0)
\end{equation}
where 
\begin{equation}
    M=4A+2\gamma\left(\Delta_{1}-\mathds1\right)
\end{equation}

The above equations give a slightly different perspective on the edge
modes, namely as eigenmodes of $M$. If we say that $v$ is a edge eigenmode
of $M$ with (complex) eigenvalue $\lambda_{v}$, whereas all the
bulk modes are indexed by $k$. Then this will clearly show up as
the decay time of the 2-time correlation function because the evolution
with respect to $\Delta t$ follows this same Heisenberg evolution:
\begin{equation}
C_{ij}(t,\Delta t)=\mathrm{Tr}\left[\eta_{i}(\Delta t)\eta_{j}\rho(t)\right]\implies\frac{dC_{ij}}{d\Delta t}=\left(MC\right)_{ij}
\end{equation}
In particular, if $\eta_{i}(0)=\alpha_{v}\eta_{v}+\sum_{k}\alpha_{k}\eta_{k}$,
where $\eta_{v}=v\cdot\eta$ is the edge operator, then
\begin{align}
\eta_{i}(\Delta t) & =\alpha_{v}\eta_{v}e^{\lambda_{v}\Delta t}\\
&+\sum_{k}\alpha_{k}\eta_{k}e^{\lambda_{k}\Delta t}\stackrel{\Delta t\to\infty}{\longrightarrow}\alpha_{v}\eta_{v}e^{\lambda_{v}\Delta t}\\
\implies C_{ij}(t,\Delta t) & \stackrel{\Delta t\to\infty}{\longrightarrow}\alpha_{v}C_{ij}(t,0)e^{\lambda_{v}\Delta t}
\end{align}
assuming the edge mode is the slowest-decaying operator. 
We have checked and confirmed that the edge mode solutions of $M$ correspond exactly to the edge modes solution of the non-Hermitian Hamiltonian.

\subsection{Analytical solution for edge modes \label{sec:supp_edge_modes_M}}

Here we calculate the edge modes of matrix (with $J=1$)

\begin{equation}
M=\begin{pmatrix}0 & -2h\\
2h & -2\gamma & -2\\
 & 2 & -2\gamma & -2h\\
 &  & 2h & -2\gamma\\
 &  &  &  & \ddots
\end{pmatrix}
\end{equation}
We consider the following (unnormalized) ansatz for the edge mode
\begin{equation}
\eta_{\text{edge}}=\sum_{j=1}^{N}\left(r^{j-1}\eta_{2j-1}+Ar^{j-1}\eta_{2j}\right)=v^{T}\eta,
\end{equation}
where 
\begin{equation}
v^{T}=\begin{pmatrix}1 & A & r & Ar & r^{2} & \dots\end{pmatrix}
\end{equation}
is the vector of coefficients and
\begin{equation}
\eta^{T} \equiv \begin{pmatrix}\eta_1 & \eta_2 & \eta_3 & \eta_4 & \eta_5 & \dots\end{pmatrix}
\end{equation}
is the vector of Majorana operators. Then
\begin{align}
M\eta_{\text{edge}} & =v^{T}M\eta=Ev^{T}\eta\implies M^{T}v=\lambda v\\
\begin{pmatrix}0 & 2h\\
-2h & -2\gamma & 2\\
 & -2 & -2\gamma & 2h\\
 &  & -2h & -2\gamma\\
 &  &  &  & \ddots
\end{pmatrix}\begin{pmatrix}1\\
A\\
r\\
Ar\\
\vdots
\end{pmatrix} & =\lambda\begin{pmatrix}1\\
A\\
r\\
Ar\\
\vdots
\end{pmatrix}.
\end{align}
This gives three independent equations:
\begin{align}
2Ah & =\lambda\label{eq:first_eq}\\
-2h-2\gamma A+2r & =\lambda A\label{eq:second_eq}\\
-2A-2\gamma r+2hAr & =\lambda r.\label{eq:third_eq}
\end{align}
Using Equation \ref{eq:first_eq} in \ref{eq:third_eq}, we can cancel
terms and simplify to get
\begin{align}
A & =-\gamma r\\
\lambda & =2Ah=-2h\gamma r\\
\implies-\cancel{2}h-\cancel{2}\gamma\left(-\gamma r\right)+\cancel{2}r & =\left(-\cancel{2}h\gamma r\right)\left(-\gamma r\right)\\
h\gamma^{2}r^{2}-\left(\gamma^{2}+1\right)r+h & =0\\
r & =\frac{\left(\gamma^{2}+1\right)\pm\sqrt{\left(\gamma^{2}+1\right)^{2}-4h^{2}\gamma^{2}}}{2h\gamma^{2}}
\end{align}

Let's examine a few transitions in this value. For $h=J=1$, we have
\begin{align}
\underline{h=J}:\;r & =\frac{\left(\gamma^{2}+1\right)\pm\sqrt{\left(\gamma^{2}+1\right)^{2}-4\gamma^{2}}}{2\gamma^{2}}\\
 & =\frac{\left(\gamma^{2}+1\right)\pm\left(\gamma^{2}-1\right)}{2\gamma^{2}}\\
 & =\begin{cases}
1\\
\frac{1}{\gamma^{2}}
\end{cases}
\end{align}
The $r=1$ solution corresponds to the $\xi\to\infty$ bulk phase
transition -- and associated edge mode -- which happens for arbitrary
$\gamma$ because the bulk is unaffected by $\gamma$. For $r=1$,
the energy is $\lambda=-2\gamma$, which is precisely the same real
part as the bulk energy, as expected. Meanwhile, for $\gamma>1$,
the additional solution $r=1/\gamma^{2}<1$ exists, which corresponds
to the well-localized edge mode in the FM phase. This has energy $\lambda=-2/\gamma$
which decays slower than the mode at $-2\gamma$, meaning that this
second edge mode is dominant and the additional edge mode that develops
at $h=J$ due to the bulk transition is not seen in late time dynamics.

A second transition happens at $\gamma=J=1$:
\begin{align}
\underline{\gamma=J}:\;r & =\frac{1\pm\sqrt{1-h^{2}}}{h}
\end{align}
For $h<1$ (FM), this gives one non-trivial solution, corresponding
to the $-$ root; this is again the continuation of the FM edge mode.
For $h>1$, the square root becomes strictly imaginary. Then
\begin{equation}
\left|r\right|^{2}=\frac{1+\left(h^{2}-1\right)}{h^{2}}=1
\end{equation}
meaning we again have a $\xi\to\infty$ transition, but now with $\text{arg}\left(r\right)\neq0$.
In particular, as $h\to\infty$, we have $r\to\pm i$ suggesting that
the edge mode transition is dominated by bulk modes with $k=\pm\pi/2$.

Finally, we have the exceptional point (a.k.a. over/under-damped)
transition that occurs when the square root in $r$ goes through zero,
causing solutions to $r$ (and therefore $\lambda$) to go from purely
real to complex. This happens when
\begin{equation}
\left(\gamma^{2}+1\right)^{2}-4h^{2}\gamma^{2}=0\implies h=\frac{\gamma^{2}+1}{2\gamma}.
\end{equation}
Note that this transition only occurs for $\gamma>1$ to satisfy the
requirement that $\left|r\right|\leq1$.

\section{Third quantization}

An alternative route to solving open quantum systems involves promoting the density matrix to a ``supervector'' and the time evolution function (Liouvillian) to a non-Hermitian ``superoperator.'' This is is commonly referred to as third quantization. In this section, we discuss how our model can be solved using third quantization and spell out the explicit connection to second-quantized Heisenberg time evolution, as was used in the previous section.

\subsection{Dynamics and edge modes from third quantization \label{sec:supp_third_quantization}}

Generally simulations of such an open quantum system are challenging, particularly with symmetry-breaking field. However, by vectorizing the density matrix, one can map the Liouvillian to a non-Hermitian Hamiltonian acting on free Majorana fermions. The non-Hermitian Hamiltonian has the ladder form (see Fig.~\ref{fig:model}b): 
\begin{equation}
\mathcal{L}\cong -i H \otimes   \noindent
\mathds{1} +i   \noindent
\mathds{1}
 \otimes H^T+ \gamma (\sigma_{1} ^z \otimes \sigma_{1} ^z - \mathds{1} \otimes \mathds{1)}=-i\mathcal{H}
\end{equation}
\mkaddcomment{Combine equations}
Our non-Hermitian Hamiltonian on a ladder system takes the form
\begin{equation}
\begin{split}
    \mathcal{H}=&-J\sum_{n=1} ^{L-1} \sigma_{n} ^z \sigma_{n+1} ^z - h\sum_{n=1} ^{L} \sigma_n ^x +J\sum_{n=1} ^{L-1} \tau_{n} ^z \tau_{n+1} ^z\\
    &+h\sum_{n=1} ^{L} \tau_n ^x+i \gamma \sigma_1 ^z \tau_1 ^z
\end{split}
\end{equation}
Noting that the dissipation maps to a complex Ising interaction, this Hamiltonian can clearly be mapped to a model of free, hopping Majoranas. Specifically,  we use Jordan-Wigner transformation 
\begin{equation}
    \mkaddcomment{[write J-W here]}
\end{equation}
to write our non-Hermitian Hamiltonian as
\begin{equation}
\begin{split}
    \mathcal{H}  &=ih\sum_{n=1} ^{L-1}\eta_{B,n} \eta_{A,n+1} +iJ \sum_{n=1} ^{L} \eta_{A,n} \eta_{B,n}\\
    &-ih\sum_{n=1} ^{L-1}\eta' _{B,n} \eta' _{A,n+1} -iJ \sum_{n=1} ^{L} \eta' _{A,n} \eta' _{B,n} -\gamma \eta' _{A,1} \eta_{A,1}
\end{split}
\end{equation}
where $\eta_{A,n}(\eta_{B,n})$ and $\eta' _{A,n}(\eta' _{B,n})$ are free Majorana fermions at the lower and upper legs of the ladder (Fig.~\ref{fig:model}) respectively.

\mkaddcomment{I woudl probably just summarize here and remove the rest of this section because we prove later that this non-Hermitian model from third quantization is equivalent to the dynamics Majorana matrix from second quantization and edge mode solution to the latter is cleaner.}

In fact, the appearance of this relationship between edge modes and boundary spin dynamics allows us to make analytical predictions for these properties. We can in fact solve for the edge modes exactly, starting by postulating edge states of the following form:
\begin{equation}
    \tilde{c}=\sum_{n=1} ^{L} (\alpha_n \eta_{A,n}+\beta_n \eta_{B,n}) +\sum_{n=1} ^{L} (\alpha' _n \eta' _{A,n}+\beta' _n \eta' _{B,n})
\end{equation}
where $\alpha_n=r^{n-1}$, $\beta_n=B r^{n-1}$, $\alpha' _n=A' r^{n-1}$, $\beta' _n=B'  r^{n-1}$ and $|r|\leq 1$ is required to get well defined edge state.
In order to solve for the exact form of edge states we first calculate the following commutation relation \mkaddcomment{Cut some of this text/move to supplement?}
\begin{equation}
    [H,\tilde{c}]=E\tilde{c}
\end{equation}
After calculating the commutator and equating the coefficients for each Majorana fermion we get the following equations
\begin{align*}
    E \alpha_1 = & 2 \gamma \alpha' _1 + 2 i h \beta_1\\
    E \alpha_{n>1}=&2 i h \beta_j -2i \beta_{n-1}\\
    E \beta_{n}=&2 i \alpha_{n+1} -2 i h \alpha_n\\
    E \alpha' _1 = & -2 \gamma \alpha _1 - 2 i h \beta' _1\\
    E \alpha' _{n>1}=& -2 i h \beta' _n +2i \beta' _{n-1}\\
    E \beta' _{n}=&-2i \alpha' _{n+1} +2 i h \alpha' _n
\end{align*}
By solving the above equations, we can obtain the exact form of edge states and compare them to the phase diagram found from $\tau^{-1}$. We find that there are zero edge modes for paramagnetic phase, two edge modes for ferromagnetic phase and four edge modes for phase III. 

Below are the edge modes solution for ferromagnetic phase \mkaddcomment{cut? adjust?}
\begin{equation}
\begin{split}
    \tilde{c}& =\mathcal{N}\Bigg( \sum_{n=1} ^{L} ( \eta_{A,n}+ B \eta_{B,n})r^{n-1} \\
    &+\sum_{n=1} ^{L} (A'_n \eta'_{A,n}+B'  \eta'_{B,n})r^{n-1} \Bigg)
\end{split}
\end{equation}
where $\mathcal{N}$ is a normalization constant and
\begin{align*}
    r=& \frac{1+\gamma^2-\sqrt{(1+\gamma^2)^2-4h^2\gamma^2}}{2h\gamma^2}\\
    B=&\mp \gamma r\\
    A'=&\pm i\\
    B'=&i \gamma r
    \end{align*}
\begin{equation}
    E_{1,2}=
    -i(\gamma\pm \sqrt{2}\sqrt{\frac{1-2 h^2 \gamma^2 +\gamma^4  +(\gamma^2-1)\sqrt{(1+\gamma^2)^2-4 h^2 \gamma^2}}{\gamma^2}})
\end{equation}
at the phase boundary between ferromagnetic phase and paramagnetic phase $(h=1,\gamma<0)$ $r=1$  (correlation length diverges) indicating second order phase transition.

Edge modes for Phase-III

\begin{align*}
    \tilde{c}=&\mathcal{N}\Bigg(\sum_{n=1} ^{L} ( \eta_{A,n}+ B \eta_{B,n})r^{n-1} +\sum_{n=1} ^{L} (A' _n \eta' _{A,n}\\
    &+B'  \eta' _{B,n})r^{n-1}\Bigg)
\end{align*}
where
\begin{align*}
    r=& \frac{1+\gamma^2 \mp \sqrt{(1+\gamma^2)^2-4h^2\gamma^2}}{2h\gamma^2}\\
    B=&\mp \gamma r\\
    A'=&\pm i\\
    B'=&i \gamma r\\
    E_{1,2}=&-i ( \gamma  - \frac{\gamma^2-1}{\gamma}) \pm\frac{\sqrt{4 h^2 \gamma^2 -(1+\gamma^2)^2}}{\gamma}  \\
    E_{3,4}=& -i (\gamma  + \frac{\gamma^2-1}{\gamma}) \pm\frac{\sqrt{4 h^2 \gamma^2 -(1+\gamma^2)^2}}{\gamma} 
\end{align*}
at the phase boundary $(h=\frac{\gamma^2+1}{2\gamma})$ between ferromagnetic phase and Phase-III  $r=0.5$ (finite correlation length) indicating first order phase transition between two phases. Similarly at the phase boundary between Phase-III and paramagnetic phase $(h\geq 1, \gamma=1)$ $r$ is a complex number with magnitude $|r=1|$ (but with finite correlation length) indicating first order phase transition between the two phases.

\subsection{Relationship between third-quantized Hamiltonian and second-quantized Heisenberg evolution \label{sec:supp_connecting_second_and_third_quantization}}

We claimed earlier that the non-Hermitian Hamiltonian in the doubled Hilbert space has identical information as the time-dependent correlation function, determined by the matrices $M$ and $\mathcal{M}$. \hl{[check notation on these, probably want to adjust.]} We now prove that they are directly related by a unitary rotation.

Specifically, let us start by writing the third-quantized Hamiltonian as
\begin{equation}
\begin{split}
    \mathcal{H}  &=ih\sum_{n=1} ^{L-1}\eta_{B,n} \eta_{A,n+1} +iJ \sum_{n=1} ^{L} \eta_{A,n} \eta_{B,n}\\
    &-ih\sum_{n=1} ^{L-1}\eta' _{B,n} \eta' _{A,n+1} -iJ \sum_{n=1} ^{L} \eta' _{A,n} \eta' _{B,n} -\gamma \eta' _{A,1} \eta_{A,1} \\ 
    = i \eta^T \mathcal{M} \eta
\end{split}
\end{equation}
where $\mathcal M$  is chosen to be an antisymmetric matrix.

We believe that these matrices capture the same information and so
are -- in some sense that we should now try to figure out -- isospectral.
The obvious thing is to try to diagonalize the $\gamma$ term, since
that is the big difference between the matrices. This suggests a unitary
transformation of $\tilde{M}$ given by
\begin{align}
\mathcal{V} & =\frac{1}{\sqrt{2}}\left(\begin{array}{cccccccccc}
1 &  &  &  &  &  &  &  &  & i\\
 & 1 &  &  &  &  &  &  & i\\
 &  & 1 &  &  &  &  & i\\
 &  &  & \ddots &  &  & \iddots\\
 &  &  &  & 1 & i\\
 &  &  &  & i & 1\\
 &  &  & \iddots &  &  & \ddots\\
 &  & i &  &  &  &  & 1\\
 & i &  &  &  &  &  &  & 1\\
i &  &  &  &  &  &  &  &  & 1
\end{array}\right)\equiv\frac{1}{\sqrt{2}}\left(\mathds{1}+i\mathcal{R}\right)\\
\mathcal{M}_{V} & =\mathcal{V}^{\dagger}\mathcal{M}\mathcal{V}\\
 & =\frac{1}{2}\left(\mathcal{M}+i\left[\mathcal{M},\mathcal{R}\right]+\mathcal{R}\mathcal{M}\mathcal{R}\right)\\
\left[\mathcal{M},\mathcal{R}\right] & =\frac{1}{2}\left(\begin{array}{cccccc}
0 & -h\\
h & \ddots\\
 &  & 0 & i\gamma\\
 &  & -i\gamma & 0\\
 &  &  &  & \ddots & h\\
 &  &  &  & -h & 0
\end{array}\right)\left(\begin{array}{cccccc}
 &  &  &  &  & 1\\
 &  &  &  & 1\\
 &  &  & 1\\
 &  & 1\\
 & 1\\
1
\end{array}\right)-\frac{1}{2}\left(\begin{array}{cccccc}
 &  &  &  &  & 1\\
 &  &  &  & 1\\
 &  &  & 1\\
 &  & 1\\
 & 1\\
1
\end{array}\right)\left(\begin{array}{cccccc}
0 & -h\\
h & \ddots\\
 &  & 0 & i\gamma\\
 &  & -i\gamma & 0\\
 &  &  &  & \ddots & h\\
 &  &  &  & -h & 0
\end{array}\right)\\
 & =\frac{1}{2}\left(\begin{array}{cccccc}
 &  &  &  & -h & 0\\
 &  &  &  & \iddots & h\\
 &  & i\gamma & 0\\
 &  & 0 & -i\gamma\\
h & \iddots\\
0 & -h
\end{array}\right)-\frac{1}{2}\left(\begin{array}{cccccc}
 &  &  &  & -h & 0\\
 &  &  &  & \iddots & h\\
 &  & -i\gamma & 0\\
 &  & 0 & i\gamma\\
h & \iddots\\
0 & -h
\end{array}\right)=\left(\begin{array}{cccccc}
0\\
 & \ddots\\
 &  & i\gamma & 0\\
 &  & 0 & -i\gamma\\
 &  &  &  & \ddots\\
 &  &  &  &  & 0
\end{array}\right)\\
\mathcal{R}\mathcal{M}\mathcal{R} & =\frac{1}{2}\left(\begin{array}{cccccc}
0 & -h\\
h & \ddots\\
 &  & 0 & -i\gamma\\
 &  & i\gamma & 0\\
 &  &  &  & \ddots & h\\
 &  &  &  & -h & 0
\end{array}\right)\\
\mathcal{M}_{V} & =\frac{1}{2}\left[\left(\begin{array}{cccccc}
0 & -h\\
h & \ddots\\
 &  & 0 & 0\\
 &  & 0 & 0\\
 &  &  &  & \ddots & h\\
 &  &  &  & -h & 0
\end{array}\right)+i\left(\begin{array}{cccccc}
0\\
 & \ddots\\
 &  & i\gamma & 0\\
 &  & 0 & -i\gamma\\
 &  &  &  & \ddots\\
 &  &  &  &  & 0
\end{array}\right)\right]=\frac{1}{2}\left(\begin{array}{cccccc}
0 & -h\\
h & \ddots\\
 &  & -\gamma & 0\\
 &  & 0 & \gamma\\
 &  &  &  & \ddots & h\\
 &  &  &  & -h & 0
\end{array}\right)
\end{align}
which rotates the forward and backwards Majoranas into their $y$-basis:
$\eta_{j}\pm i\eta_{j}^{\prime}$. This unitary has block-diagonalized
the matrix. In fact, notice that the lower-right block is equal to
$-\tilde{M}/4$, where $\tilde{M}$ is the matrix that determines
equal-time correlations. By inverting the rows and columns of the
upper left block via further unitary matrix $\mathcal{R}_{U}$, we
can relate that block to the non-equal-time generating matrix $M$:
\begin{align}
\mathcal{R}_{U} & =\left(\begin{array}{cccccc}
0 & 0 & 1\\
0 & \iddots & 0\\
1 & 0 & 0\\
 &  &  & 1 & 0 & 0\\
 &  &  & 0 & \ddots & 0\\
 &  &  & 0 & 0 & 1
\end{array}\right)\\
\mathcal{M}_{f} & =\mathcal{R}_{U}\mathcal{M}_{V}\mathcal{R}_{U}\\
 & =\frac{1}{2}\left(\begin{array}{cccccccc}
-\gamma & h\\
-h & \ddots & J\\
 & -J & 0 & h\\
 &  & -h & 0\\
 &  &  &  & \gamma & h\\
 &  &  &  & -h & \ddots & J\\
 &  &  &  &  & -J & 0 & h\\
 &  &  &  &  &  & -h & 0
\end{array}\right)=\left(\begin{array}{cc}
-M/4 & 0\\
0 & -\tilde{M}/4
\end{array}\right).
\end{align}
\hl{[adjust the notation so that these factor of "-" and, perhaps, 1/4, are gone.}

\section{Equilibrium observables}

In this section we treat two quasi-equilibrium observables in the non-equilibrium steady state: energy current from the bath to the system and magnetic susceptibility. We show numerically and analytically that neither has measurable singularities at the locations where edge state phase transitions occur in the autocorrelation function.

\subsection{Energy Current \label{sec:supp_energy_current}}

As we have seen in the previous section of two-time correlations and edge state, singularities appear at the edge of the boundary-dissipation Ising chain. However, it would be interesting to look at other observables where this singularity might show up, and that are easier to measure. One option is an energy current from the $T=\infty$ bath. Here we show how to calculate this energy current.

\begin{figure}[h]
    \centering
    \includegraphics[width=1.0\columnwidth]{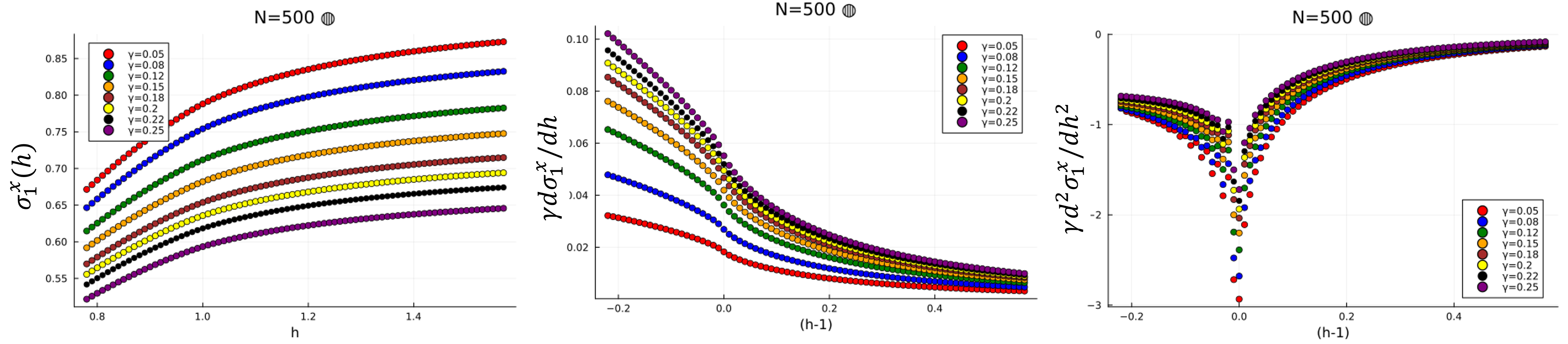}
    \caption{a) Boundary magnetization (which is proportional to energy current Eq.~\ref{IE}) for different values of $\gamma$ b) First derivative of $\sigma_1 ^x $ as function of h, c) Second derivative $\sigma_1 ^x $ as function of h.}
       \label{fig:energycurrent}
\end{figure}

The idea is that energy taken from bath is equal to energy absorbed by system. So, we have
\begin{align}
I_{E} & =\frac{d\langle H\rangle}{dt}\\
 & =\frac{d\text{Tr}\left(\rho H\right)}{dt}\\
 & =\text{Tr}\left(\dot{\rho}H\right)\\
 & =\text{Tr}\left[\left(L\rho L^{\dagger}-\frac{1}{2}\left\{ L^{\dagger}L,\rho\right\} \right)H\right]
\end{align}
Note that the Hamiltonian part of time evolution conserves energy, so drops out. For our Lindbladian, this is
\begin{align}
I_{E} & =\gamma\text{Tr}\left[\left(\sigma_1 ^z \rho \sigma_1 ^z -\rho\right)H\right]\\
 & =\gamma\text{Tr}\left[\rho\left(\sigma_1 ^z H \sigma_1 ^z-H\right)\right]\\
 & =\gamma\text{Tr}\left[\rho\left(-2h \sigma_1 ^x\right)\right]\\
 & =-2h\gamma\langle \sigma_1 ^z \rangle
\label{IE}
\end{align}
So it's just proportional to the static transverse magnetization at the boundary. 
As shown in Fig.~\ref{fig:energycurrent}, we don't see the singular behavior in energy current, which is quite unexpected based on our initial intuition that there are dissipative edge modes at the dissipative boundary and we would see their effect in energy current. However, based on our data, it appears that their influence on current flow is rounded off due to the bulk eigenmodes.

\begin{figure}[h!]
\centering
\includegraphics[width=0.3\linewidth]{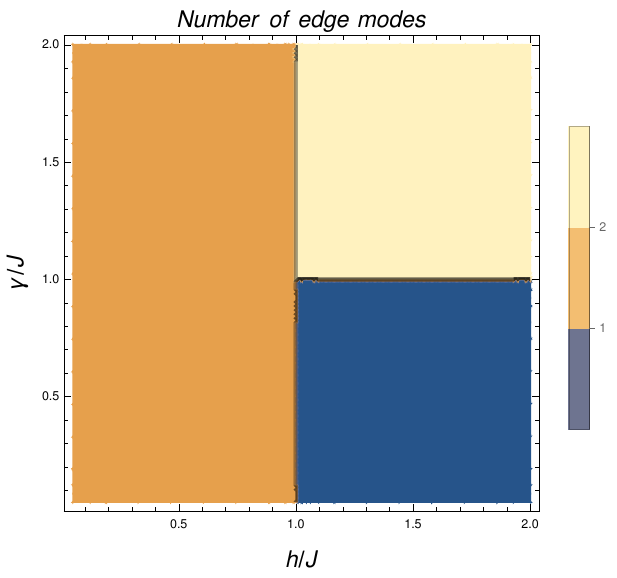}
\includegraphics[width=0.67\linewidth]{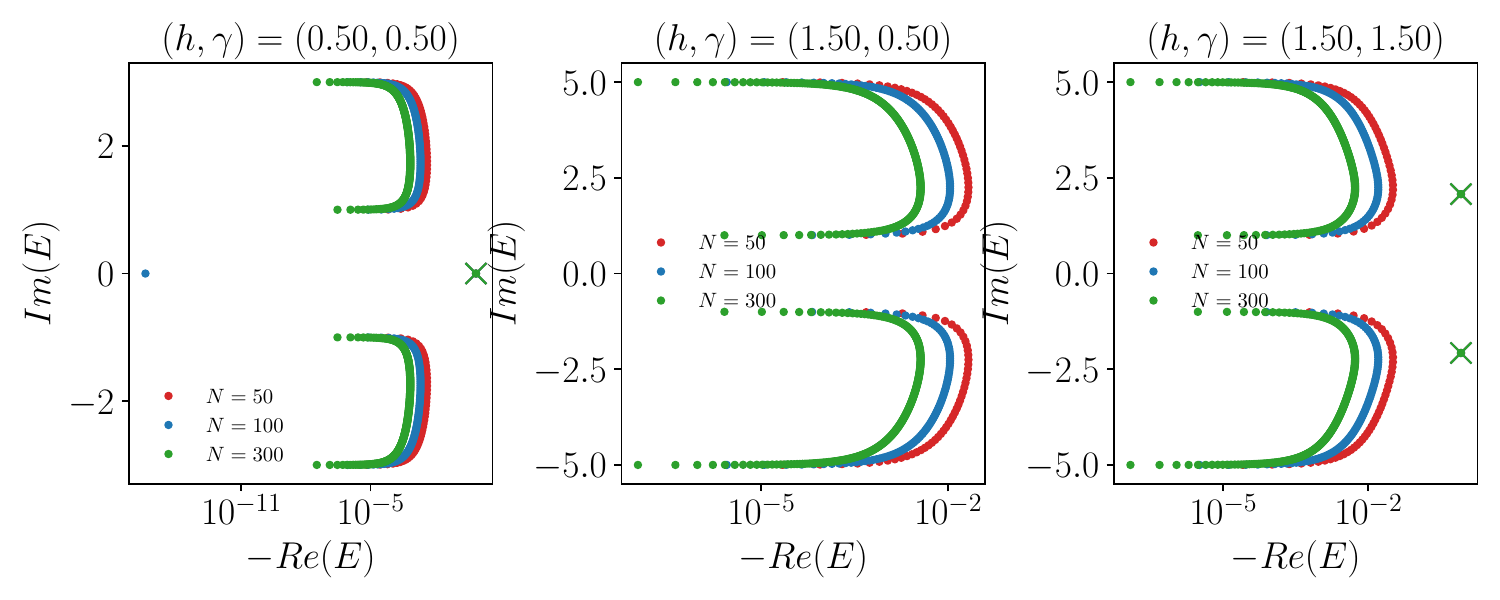}
\caption{(a) Number of edge modes of $\tilde{M}$ (cf. Eq.~\eqref{eq_Mtilde}) with $|r|<1$ from Eq.~\eqref{eq_solution_edge_Mtilde} (blue = 0, orange = 1, yellow = 2) . (b-d) full spectrum of $\tilde{M}$. The crosses indicates the dissipative edge modes found analytically. In the ferromagnetic phase we have zero-energy edge mode localized around the right edge of the system (which is weakly touched by $\gamma$, and for $N\to \infty$ it is ``essentially'' infinitely long lived). The left edge is instead with a finite lifetime, and it results to be the fastest deaying mode (generally, except when Zeno effect takes over).}
\label{fig_number_of_edges}
\end{figure}

\begin{figure}[h!]
\centering
\includegraphics[width=0.67\linewidth]{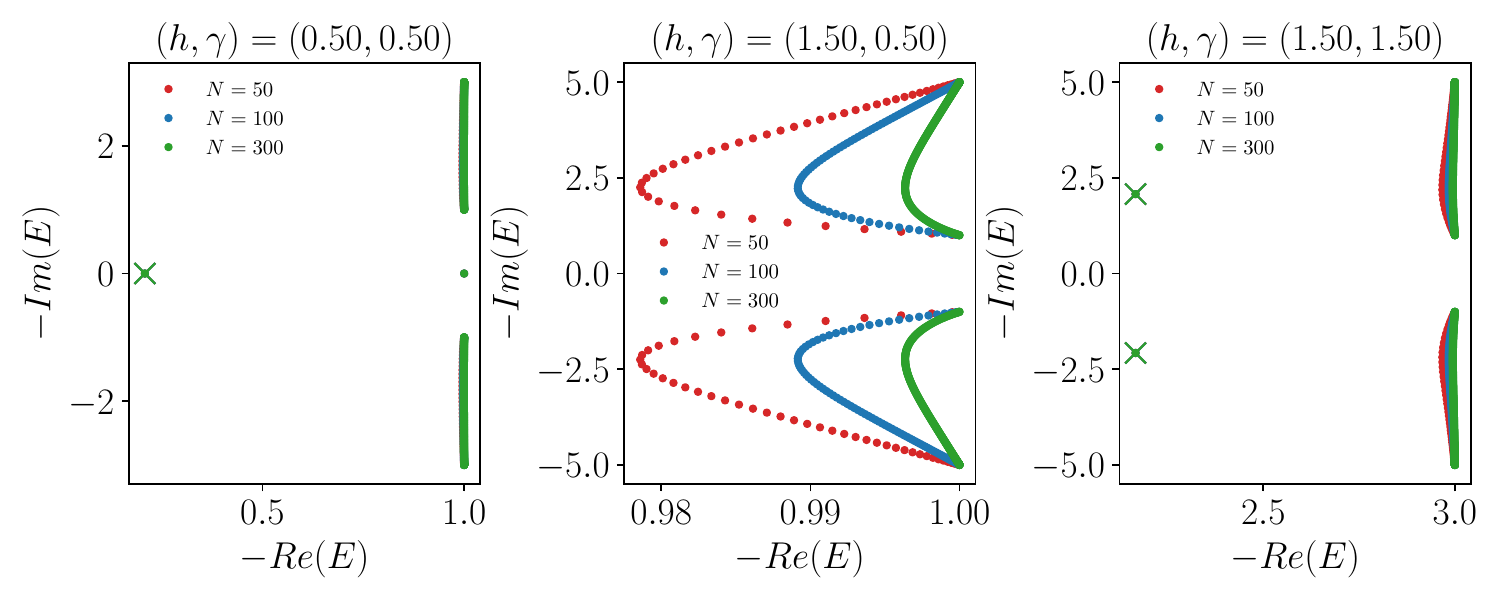}
\caption{Spectrum and edge modes of the matrix $M = 4A + 2\gamma(\Delta_1 - \bm{1})$, controlling the evolution of the autocorrelation function. Notice that edge modes are the slowest-decaying ones, contrary to what occurs in the equal-time correlator (see Fig.~\ref{fig_number_of_edges}).}
\label{fig_spectrum_M}
\end{figure}

We want to see whether the singularities in the autocorrelation function reflects on equal time observables. Adopting the convention of Mike, Umar et al, we write the single particle Hamiltonian $\mathcal{H}_{MF}=iA$, wiht $A$ antisymmetric and real.
Given $C_{ij}(t) = \langle \eta_i(t)\eta_j(t)\rangle$, their dynamics is governed by
\begin{equation}
\label{eq_motion_C}
\frac{d\bm{C}}{dt}=4[A,C]-2\gamma\{\Delta_1,\bm{C}\} + 4 \gamma \Delta_1 \bm{C}\Delta_1.
\end{equation}
where $\bm{C}$ is a matrix and, analogusly to what can be done for density matrices.\footnote{
\textbf{Considerations about past notes.} In those notes, it is claimed from Eq.~(9) to Eq.~(12) (numbering refers to the notes) that the dynamics of equal time correlator is governed by a shifted matrix $M$, specifically by $M'=M-2\gamma \bm{1}$. This is incorrect, since that claim was based on a definition of $M$ with the wrong sign assigned to the dissipative part ($\gamma \to -\gamma$).}
The equation of motion of the equal time correlator can be linked to the matrix
\begin{equation}
\label{eq_Mtilde}
\tilde{M}=4A-2\gamma \Delta_1,
\end{equation}
where the $2N \times 2N$ matrix $A$ is the single-particle Hamiltonian in the Majorana fermions
\begin{equation}
A = 
\begin{pmatrix}
0 & -h/2 &  &  \\ 
h/2 & 0 & -J/2 &  \\ 
 & J/2 & 0 & -h/2 \\ 
 &  & h/2 & 0
\end{pmatrix} 
\end{equation}
Indeed, we can write (see Mike's notes on eigenvectors)
\begin{equation}
\begin{split}
\frac{d\bm{C}}{dt}&=\tilde{M}\bm{C}+ \bm{C}\tilde{M}^\dagger\\
&=\left(4A-2\gamma \Delta_1\right) \bm{C} + \bm{C}\left(-4A - 2\gamma\Delta_1\right)\\
&=4[A,\bm{C}] - 2\gamma \{\Delta_1,\bm{C}\},
\end{split}
\end{equation}
which is equal to the equation of motion in Eq.~\eqref{eq_motion_C}, except for the jump terms. From Mike's notes, the dynamics of $\bm{C}$ depends on the spectral properties of $\tilde{M}$ (this is straightforward adopting a superoperator formalism). We aim to see if such matrix admits edge modes, which could have an impact in the dynamics analogously to what occurs in the out-of-time correlator. To this end, we look for eigenstates of the form of edge modes
\begin{equation}
\tilde{v} = \sum_{j=1}^N \left(ar^{j-1} \eta_{2j-1} + b r^{j-1}\eta_{2j}\right)
\end{equation}
with $a$, $b$ and $r$ numbers. We assume that edge modes have energy $E$ (not necessarily zero). Moreover, since we are dealing with dissipative system we do not expect real energies. Notice that $A$ has a completely immaginary spectrum due to its properties (which leads to real eigenvalues of the single-particle Hamiltonian $\mathcal{H} = i A$), but due to $\Delta_1$ we expect a real part, which would lead to decay.
We solve
\begin{equation}
\begin{pmatrix}
-2\gamma & -2h & • & • \\ 
2h & 0 & -2J & • \\ 
• & 2J & 0 & -2h \\ 
• & • & 2h & 0
\end{pmatrix} 
\begin{pmatrix}
a \\ 
b \\ 
ar \\ 
br \\ 
ar^2 \\ 
br^2 \\ 
ar^3 \\ 
\vdots
\end{pmatrix} 
=
E
\begin{pmatrix}
a \\ 
b \\ 
ar \\ 
br \\ 
ar^2 \\ 
br^2 \\ 
ar^3 \\ 
\vdots
\end{pmatrix} 
\end{equation}
We obtain a set of coupled equations, which are all linearly dependent to these three equations:
\begin{equation}
\left\{
\begin{split}
&-2\gamma a - 2bh = Ea\\
&2ha - 2J r a = Eb\\
&2Jb - 2h b r = E a r
\end{split}\right.
\end{equation}
We set $J=1$ (freedom in setting the energy scale) and $a=1$ (we look for solutions of this kind, since $a$ can be always absorbed in $b$. Nb: $a=0$ leads to null solution, which is unphysical). We obtain a linear system with two solutions
\begin{equation}
\label{eq_solution_edge_Mtilde}
\begin{split}
E_{1,2} &= \frac{1-\gamma ^2 \pm \sqrt{\gamma ^4+\gamma ^2 \left(2-4 h^2\right)+1}}{\gamma }\\
b_{1,2} &= -\frac{1+\gamma ^2 \pm \sqrt{\gamma ^4+\gamma ^2 \left(2-4 h^2\right)+1}}{2 \gamma h}\\	
r_{1,2} &= \frac{1+\gamma ^2 \pm \sqrt{\gamma ^4+\gamma ^2 \left(2-4 h^2\right)+1}}{2 \gamma
   ^2 h}
\end{split}
\end{equation}
We have localized edge modes for $|r|<1$. We see that the energies have always a non-zero real part. Thus, we expect decay of such edge modes. In Fig.~\ref{fig_number_of_edges} we show the number of edge modes ($|r| < 1$) which the system admits. Inspecting the full spectrum, we notice that the edge modes are the fastest decaying one, contrary to what occurs for the matrix $M$ controlling the dynamics of out-of-time-correlation function (cf. Fig.~\ref{fig_spectrum_M}) where they are the slowest decaying. Thus, we do not expect that such edges have an impact in the long-time dynamics of the magnetization. As a consequence, we do expect singularities.

\subsection{Magnetic susceptibility \label{sec:supp_magn_susc}}

\hl{***write me***}

\section{Additional data for interacting model}

In this section, we provide additional information and data on the interacting Ising chain and simulations using a variant of time-evolving block decimation (TEBD).

\subsection{Numerical fits for TEBD of non-interacting data \label{sec:tebd_non_interacting}}

\hl{***write me***}

\subsection{Alternative integrability-breaking interactions}

Here we consider the Hamiltonian
\begin{equation}
\label{eq_H2}
H_2 = - \sum_{j} z_j z_{j+1} + h_x \sum_j x_j + J_{xx}\sum_{j} x_j x_{j+1}
\end{equation}
Taking inspiration from the integrable case, we use a dressed exponential fit in the paramagnetic phase and $\gamma<1$, otherwise a purely exponential one (cf. Fig.~\ref{fig_final_plots_novel_fit_H2}).

\begin{figure}[h!]
\centering
\includegraphics[width=0.48\linewidth]{figures/TN_non_integrable_swipe_gamma_autocorrelation_h[0.5, 1.5]_Jxx[-0.01, -0.05, -0.1, -0.15]_JzzzFalse_novel_fit_final_plots.pdf}
\includegraphics[width=0.48\linewidth]{figures/TN_non_integrable_swipe_h_autocorrelation_gamma[0.5, 1.5]_Jxx[-0.01, -0.05, -0.1]_JzzzFalse_novel_fit_final_plots.pdf}
\caption{The results are given by fitting either exponential or dressed power law  (cf. Eq.~\eqref{eq_ct_power_law_dress}). We force power law fit in $h>1$ and $\gamma<1$ region.  The dashed lines in the plots $1/\tau$ are the results from the integrable case. We observe abrupt jump, which is absent if instead we fix the fitting procedure. Could it be due to the fact that the transition in $\gamma$ does not occur anymore at $\gamma=1$, but it is shifted?}
\label{fig_final_plots_novel_fit_H2}
\end{figure}

\bibliography{references.bib}
